Sensitivity of Biosignatures on Earth-like Planets orbiting in the Habitable Zone of
Cool M-Dwarf Stars to varying Stellar UV Radiation and Surface Biomass Emissions

*Submitted to the Planetary and Space Science (PSS) Special Issue "Planetary Evolution and Life"*

J. L. Grenfell[1#], S. Gebauer[1], P. v. Paris[2], M. Godolt[1,2] and H. Rauer[1,2]

[1] Zentrum für Astronomie und Astrophysik, Technische Universität Berlin (TUB), Hardenbergstr. 36, 10623 Berlin, Germany, email: lee.grenfell@dlr.de

[2] Institut für Planetenforschung, Deutsches Zentrum für Luft- und Raumfahrt (DLR), Rutherford Str. 2, 12489 Berlin, Germany

#Corresponding author, contact details:

Email: lee.grenfell@dlr.de

Tel: +49 30 314-25463

FAX: +49 30 314-24885

**Abstract**: *We find that variations in the UV emissions of cool M-dwarf stars have a potentially large impact upon atmospheric biosignatures in simulations of Earth-like exoplanets i.e. planets with Earth's development, and biomass and a molecular nitrogen-oxygen dominated atmosphere. Starting with an assumed black-body stellar emission for an M7 class dwarf star, the stellar UV irradiation was increased stepwise and the resulting climate-photochemical response of the planetary atmosphere was calculated. Results suggest a "Goldilocks" effect with respect to the spectral detection of ozone. At weak UV levels, the ozone column was weak (due to weaker production from the Chapman mechanism) hence its spectral detection was challenging. At strong UV levels, ozone formation is stronger but its associated stratospheric heating leads to a weakening in temperature gradients between the stratosphere and troposphere, which results in weakened spectral bands. Also, increased UV levels can lead to enhanced abundances of hydrogen oxides which oppose the ozone formation effect. At intermediate UV (i.e. with x10 the stellar UV radiative flux of black body Planck curves corresponding to spectral class M7) the conditions are "just right" for spectral detection. Results suggest that the planetary $O_3$ profile is sensitive to the UV output of the star from ~(200-350) nm. We also investigated the effect of increasing the top-of-atmosphere incoming Lyman-$\alpha$ radiation but this had only a minimal effect on the biosignatures since it was efficiently absorbed in the uppermost planetary atmospheric layer, mainly by abundant methane. Earlier studies have suggested that the planetary methane is an important stratospheric heater which critically affects the vertical temperature gradient, hence the strength of spectral emission bands. We therefore varied methane and nitrous oxide biomass emissions, finding e.g. that a lowering in methane emissions by x100 compared with the Earth can influence temperature hence have a significant effect on biosignature spectral bands such as those of nitrous oxide. Our work emphasizes the need for future missions to characterize the UV of cool M-dwarf stars in order to understand potential biosignature signals.*

## 1. Introduction

Finding evidence of life on an exoplanet is a fundamental goal of exoplanet science. Planets orbiting in the Habitable Zone (HZ) of cool M-dwarfs are potentially favoured targets (Scalo et al., 2007) due to their low planet to star flux contrast ratios, but there are many proposed properties of these systems which could impact their habitability e.g. regarding their possible tidal-locking (e.g. Kasting et al., 1993; Selsis et al., 2007), the potentially high influx of flares (e.g. Segura et al., 2010) and/or cosmic rays into the planetary atmosphere (e.g. Grenfell et al., 2012), or their ability to retain atmospheres (Lammer et al., 2007) to name but a few. Atmospheric biosignatures are species e.g. ozone ($O_3$) or nitrous oxide ($N_2O$) which, if present in suitable amounts in a planetary atmosphere may suggest the presence of life. A possible modeling approach (e.g. Segura et al., 2003) given the complications involved is to assume the Earth's development and biomass and then to allow the atmosphere to adapt e.g. to the incoming stellar spectrum. Earlier model studies (Segura et al. 2005; Segura et al., 2010; Rauer et al., 2011) applied this approach to focus on biosignature responses and spectra for Earth-like planets orbiting in the Habitable Zone (HZ) of M-dwarf stars.

Planetary atmospheric biosignature species may respond very sensitively to Ultra-Violet (UV) emissions from the central star since in these wavelengths they are either photodissociated directly or there occur photochemical reactions which affect their abundance. The cooler M-dwarf stars are statistically older and burn more slowly compared with lower spectral classes. They have more developed convection zones and possibly larger differences in UV between flaring and quiet states. Whereas M-dwarf stars with spectral classes M6 and M7 likely possess very active chromospheres (Reid and Hawley, 2005) which implies strong UV emissions, the M8 and M9 spectral classes have less active chromospheres (e.g. Kruse et al., 2010). There are only limited observations of stellar UVemissions available (mostly for active M-dwarf stars with low to mid-range spectral classes) from which one can compile spectra and calculate the planetary atmosphere e.g. for AD-Leonis (ADL) (Segura et al., 2005), for GJ-581 (von Paris et al., 2010), and recently, for nearby M-dwarf stars (France et al., 2012). Some model studies (e.g. Rauer et al., 2011; Belu et al., 2011; Kaltenegger and Traub, 2009) therefore assumed a black-body spectrum as a first approach. However, they acknowledged that some deviations from such a spectrum may occur, not only for M-dwarf chromospheres which contribute to the UVemission, but also in their photospheres in the visible and IR due to the presence of stellar absorption lines. For the cooler M-dwarf stars (spectral classes M6-M7 and above), no E(UV) observations currently exist.

In this work we concentrate on the M7 stellar class because this was a particular focus in recent modeling studies (e.g. in Rauer et al. (2011); Grenfell et al. (2013)) and are particularly favorable targets. France et al. (2013) observed stellar flux for spectral classes ~M3 to M5, suggesting rather flat spectra with no significant strong jumps in UV. For M7, although observations in the near UV (defined as in France et al, 2013) are lacking, there are theoretical grounds (see e.g. Kaler, 2011) which suggest a (possibly) strong rise in stellar near UV ($\lambda < \sim 350nm$) for ascending red-dwarf stellar classes M5→M6→M7. Dynamo theory of stellar magnetism suggests that the increased stellar rotation together with the deeper convection leads to a strengthening in the M7 chromospheres (possibly due to changes in magnetic re-connection) relative to the underlying photospheres. Although the dynamo theory has its limitations, on the above points there is general accord. Chromospheres can be very hot i.e. temperatures of several tens of thousands of degrees and emit strongly in the UV. We investigate the uncertainty in the stellar input radiation on the atmospheric biosignatures of the planet in two ways. First, we investigate uncertainty in the *slope* in input UV from 350 to 300nm. Second, we investigate the effect of increasing stellar radiation ($\lambda < 300nm$) by a factor of x10, x100 and x1000. Lyman-$\alpha$ emissions from the central star may also impact planetary biosignatures although identifying the Ly-$\alpha$ emission from M-dwarf stars is very challenging e.g. due to uncertain absorption through the Interstellar Medium (e.g. Linsky, 2011). Finally, in addition to the effect of incoming UV on biosignatures, previous model studies (Grenfell et al., 2011; Grenfell et al., 2013; Rauer et al., 2011) also noted the potential importance of climate-chemical feedbacks which depend e.g. on surface biomass emissions and stellar insolation.

In this work we build on Rauer et al. (2011) who modelled atmospheric spectra and investigated spectral signals of biosignatures and related species assuming black-body Planck curves for a range of M-dwarf spectral classes up to M7. We start with a planetary atmosphere having Earth's biomass and an incoming top-of-atmosphere (TOA) stellar irradiation corresponding to a blackbody of a cool M-dwarf star with spectral class M7, we then vary three parameters, namely (1) the UV TOA incoming stellar radiation (see Figure 1 and text above), (2) the TOA stellar Lyman-$\alpha$ flux alone and (3) the planetary surface biomass emissions, namely for methane ($CH_4$) and $N_2O$. $CH_4$ was found by Rauer et al. (2011) to strongly affect stratospheric temperatures hence the $O_3$ spectral band; $N_2O$ is varied to investigate whether it is a "good biosignature" (i.e. for which the atmospheric abundance should respond sensitively to the biomass input). We discuss these effects upon both atmospheric biosignature abundances and theoretical spectra. Section 2 gives the model description and scenarios, section 3 presents results and section 4 gives a discussion and conclusions.

## 2. Model Descriptions, Methods and About the Runs

**2.1 Atmospheric Model description** – the basic model version is described in Rauer et al. (2011) and is based on earlier versions described in Segura et al. (2003) and Kasting et al. (1993).

***The Climate Module*** is a global-mean stationary, hydrostatic column model of the atmosphere from the surface up to $6.6 \times 10^{-5}$ bar (for the Earth this corresponds to a height of about 70km) with starting composition, pressure, and temperature based on the US-standard atmosphere (1976). Radiative transfer is based on the RRTM (Rapid Radiative Transfer Module) for the thermal radiation (see Mlawer et al. 1997) which extends from 3.07-1000μm. The shortwave radiation scheme consists of 38 spectral intervals for the major absorbers extending from (237.6 nm-4.545μm) and includes Rayleigh scattering for $N_2$, $O_2$, and $CO_2$ with cross-sections based on Vardavas and Carver (1984) and assumes a solar-zenith angle of $60^o$. For the Earth control scenario, a solar spectrum based on Gueymard (2004) is employed. See section 2.3 for details of the M-dwarf spectra used. In the troposphere, convective adjustment to the moist adiabatic lapse rate is carried out and the water vapour concentrations are calculated from a relative humidity profile based on Earth observations (Manabe and Wetherald, 1967). Clouds are not included explicitly, although these are considered in a straightforward way by adjusting the surface albedo to attain the mean surface temperature of Earth (288 K) for the Earth around the Sun run (Earth control). After converging, the climate module outputs temperature, pressure, and water abundances as input for the chemistry module.

***The Chemistry Module*** was described by Pavlov and Kasting et al. (2002). The scheme includes 54 chemical species and more than 200 chemical reactions with kinetic data taken from the Jet Propulsion Laboratory (JPL) (2003) Report. The chemistry module assumes a planet with an Earth-like development, that is, with an $N_2$-$O_2$ dominated atmosphere, etc. The reaction scheme reproduces modern Earth's atmospheric composition with a focus on biomarkers (e.g., $O_3$, $N_2O$) and key greenhouse gases such as $CH_4$. The chemical module calculates the steady-state solution of the usual 1D continuity equations by an implicit Euler method. Mixing between adjacent layers is parameterized via Eddy diffusion coefficients. From a total of 54 chemical species, 34 are "long-lived," that is, their concentrations are obtained by solving the full continuity equation. Finally, three species are set to constant abundances throughout the model atmosphere, namely, $CO_2 = 3.55 \times 10^{-4}$, $O_2 = 0.21$ and $N_2 \sim 0.78$ volume mixing ratio (vmr). Remaining species are "short-lived," i.e. their abundances are calculated from the long-lived species concentrations assuming chemical steady-state. Constant surface emissions of biogenic ($CH_3Cl$, $N_2O$) and source gas ($CH_4$, CO) species based on Earth were included. Molecular hydrogen ($H_2$) was removed at the surface with a constant deposition velocity of $7.7 \times 10^{-4}$ cm s$^{-1}$. More details of the surface fluxes and boundary conditions are given e.g. in Grenfell et al. (2011). Also included are modern-day tropospheric lightning emissions of nitrogen monoxide (NO), volcanic sulphur emissions of $SO_2$ and $H_2S$, and an effusion flux of CO and O at the upper model boundary, which represents the photolysis products of $CO_2$. The chemistry module features 108 wavelength intervals with midpoints ranging from (176.2-850) nm. The sulfate aerosol parameterisation based on Segura et al. (2003) is used. Removal at the surface of long-lived species via dry and wet deposition is included via deposition velocities (for dry deposition) and Henry's law constants (for wet deposition). After converging, the chemistry module outputs greenhouse gas abundances to the climate module. This whole process, i.e. exchange of data between the two modules, repeats until the chemical abundances and the atmospheric temperature and pressure converge.

## 2.2 Theoretical Spectral Model Description

To calculate the spectral appearance of our modelled atmospheres we use the SQuIRRL (Schwarzschild Quadrature InfraRed Radiation Line-by-line) code (Schreier and Schimpf, 2001) designed to perform high-resolution radiative transfer modeling in the IR region for a spherical atmosphere (with arbitrary observation geometry, instrumental field-of-view, and spectral response function). The scheme assumes local thermodynamic equilibrium and applies a Planck function (with appropriate layer temperatures) as the source of the emission. Cloud and haze free conditions without scattering are assumed. SQuIRRL has been verified via intercomparison exercises (e.g. Melsheimer et al., 2005). Absorption coefficients are calculated with continuum corrections using molecular spectroscopic line parameters from the HITRAN 2008 (Rothman et al., 2009) database, with emission spectra calculated assuming a pencil beam at a viewing angle of $38^{\circ}$ as used, for example, in Segura et al. (2003).

## 2.3 Model Scenarios

We modelled an Earth-like planet having an ($N_2$-$O_2$) dominated atmosphere lying in the centre of the HZ i.e. receiving a total amount of stellar energy of 1366 $Wm^{-2}$. All runs unless stated otherwise assumed modern day values for Earth's biomass and source gas abundances, a fixed surface albedo=0.21 (which reproduced the modern Earth mean surface temperature of 288K for the Earth around the Sun control run) and a constant surface relative humidity of 77% (see Rauer et al., 2011 for further details).

Four input parameters were varied to investigate the influence of stellar irradiation and planetary surface biomass emissions:

*(1) Stellar UV input* – stellar input spectra were derived in a two-step procedure. In the first step, starting with Top-of-Atmosphere (TOA) M7 input stellar black-body spectrum ($T_{eff}$=2500K) taken from Rauer et al. (2011) ("M7" - run 2). In order to simulate the effect of an active chromosphere, spectral input intensities for $\lambda$ <300nm were replaced with those from the active M-dwarf star ADL spectrum (Rauer et al., 2011) scaled so that the ADL and M7 intensities are equal at 300nm ("UV-M7" - run 3) . We took the ADL stellar spectrum from the Virtual Planetary Laboratory (VPL) website (Segura et al., 2005), which is derived from observations in the visible by Pettersen and Hawley (1989), from the IUE satellite in the UV, from observations by Leggett et al. (1996) in the near-IR and based on a nextGen stellar model spectrum for wavelengths beyond 2.4 microns (Hauschildt et al., 1999). In a second step the total spectral energy flux was adjusted by an equal factor in all model wavelength intervals so that the total net intensity equaled the present-day Total Solar Irradiance (TSI) of the Earth. We chose to modify the wavelength range $\lambda$ <300nm since  because biosignatures are strongly affected by radiation in this wavelength range. Note that due to this scaling which is applied to counter the effects of higher UV fluxes, the irradiation at longer wavelengths i.e. relevant for the atmospheric temperature calculation, is slightly reduced. Furthermore, real stellar spectra may, of course, exhibit different energy distributions in this wavelength range too. In additional runs, we repeated the two-step procedure, but using $\lambda$ <300nm ADL intensities increased by factors x10, x100 and x1000 as shown in Table 1 ("x10UV-M7" - run 4, "x100UV-M7" - run 5 and "x1000UV-M7" - run 6). The simulated planets lie at 0.15 Astronomical Units (AU). Figure 1 shows the resulting input stellar UV spectra constructed as described above. We also performed a run based on the x10-UV M7 run but where we increased by a factor of x5 the UV flux in the region 300-350nm ("x5UV-M7-Slope" - run 7 (see also Figure 2).

*(2) Stellar Lyman-alpha input spectra* – starting with the TOA spectrum composed of the M7 Planck spectrum and the ADL UV fluxes we increased the Lyman-$\alpha$ flux by a factor of x100.

*(3) $CH_4$ surface biomass* – Surface emissions were reduced by a factor of x100 compared with the modern-day Earth scenario, which has 531 Tg/year (Grenfell et al., 2011). This approach represents a lower limit for an assumed Earth-like planet with methanogenic bacteria since Earth has a constant (~>1% of the total emissions) background level of non-biogenic $CH_4$ emission via geological activity (e.g. Kvenvolden and Rogers, 2005). Two runs additionally investigated the effect of increasing the $CH_4$ surface emissions by x2 and x3. The increases are rather modest due to constraints of the radiation module (Mlawer et al., 1997) and also because at higher $CH_4$ abundances, aerosols may form which are not parameterised in the model.

(4) $N_2O$ surface biomass – weak stellar UV emissions suggest slow chemical destruction of $N_2O$ in the planetary atmosphere since this species' removal from the atmosphere is favoured by UV radiation. This could mean that

even a weak abiotic $N_2O$ source may possibly lead to a large build-up in atmospheric abundances. This effect would increase the likelihood of a "false positive" detection**.** To test this, we calculated M-dwarf scenarios varying the planetary surface $N_2O$ biomass emission.

In total we performed the following fourteen runs:

Run 1 – Earth around the Sun control run from Rauer et al. (2011) **(Earth control)**
Run 2 – M7 black-body curve stellar input spectrum from Rauer et al. (2011) **(M7)**
Run 3 – as Run 2 but with increased UV intensities for λ <300nm (from the ADL stellar spectrum) **(UV-M7)**
Run 4 – as Run 3 but with incoming TOA UV (λ <300nm) in that run increased by x10 **(x10UV-M7)**
Run 5 – as Run 3 but with incoming TOA UV (λ <300nm) in that run increased by x100 **(x100UV-M7)**
Run 6 – as Run 3 but with incoming TOA UV (λ <300nm) in that run increased by x1000 **(x1000UV-M7)**
Run 7 – as Run 4 but with x5 UV stellar flux (300-350nm)  **(x10UV-M7x5-Slope)**
Run 8 –as Run 3 but with incoming TOA Lyman-α intensities increased by x100 **(x100Ly-α-UV-M7)**
Run 9 – as Run 2 but with x100 lowered surface $CH_4$ emissions **(LOW-CH₄-M7)**
Run 10 – as Run 3 but with x100 lowered surface $CH_4$ emissions **(LOW-CH₄-UV-M7)**
Run 11 – as Run 3 but with no surface $N_2O$ biomass emissions **(ABIO-N₂O-UV-M7)**
Run 12 – as Run 2 but with lowered (by x1000) $N_2O$ biomass emissions **(LOW-N₂O-M7)**
Run 13 – as Run 3 but with increased (by x2) $CH_4$ biomass emissions **(x2CH4-UV-M7)**
Run 14 – as Run 3 but with increased (by x3) $CH_4$ biomass emissions **(x3CH4-UV-M7)**

## 3. Results

### 3.1 Atmospheric Column and Temperature Responses

Table 1 shows column amounts in Dobson Units (DU) of $O_3$, $N_2O$ and $CH_4$ and surface (middle atmosphere) temperatures (K) for the fourteen runs:

Table 1: Column amounts of $O_3$, $N_2O$ and $CH_4$ in Dobson Units (DU, 1DU=2.69x10$^{16}$ molecules cm$^{-2}$), surface temperature ($T_0$) in Kelvin (K), and mid-stratospheric temperature ($T_{strat}$) defined as the model gridpoint lying closest to 50mb which corresponds to the low to mid-stratosphere for Earth.  Note that the model gridscale is variable depending on the converged pressure and temperature profiles.

| Run Number | $O_3$ (DU) | $N_2O$ (DU) | $CH_4$ (DU) | $T_0$ (K) ($T_{strat}$) |
|:---:|:---:|:---:|:---:|:---:|
| (1) (Earth control) | 305.0 | 232.7 | 1240.3 | 288.1 (216.7) |
| (2) (M7) | 32.4 | 256,499 | 4,581,971 | 292.0 (263.2) |
| (3) (UV-M7) | 44.0 | 1611 | 2,115,927 | 284.7 (257.9) |
| (4) (x10UV-M7) | 334.7 | 583.9 | 197,759 | 299.2 (231.3) |
| (5) (x100UV-M7) | 593.3 | 332.0 | 49,266 | 298.9 (228.9) |
| (6) (x1000UV-M7) | 454.0 | 211.9 | 25,539 | 288.9 (229.5) |
| (7) (x10UV-M7x5-Slope) | 229.4 | 549 | 148,154 | 298.5 (234.1) |
| (8) (x100Ly-α-UV-M7) | 44.0 | 1611 | 2.1x10$^7$ | 284.7 (257.9) |
| (9) (LOW-CH₄-M7) | 29.9 | 270,416 | 530,932 | 312.1 (241.9) |
| (10) (LOW-CH₄-UV-M7) | 144.6 | 1,880 | 30,238 | 298.1 (222.8) |
| (11) (ABIO-N₂O-UV-M7) | 42.1 | 1.3x10$^{-2}$ | 2,347,637 | 281.7 (259.3) |
| (12) (LOW-N₂O-M7) | 1.7 | 2,880 | 2.1x10$^7$ | 246.6 (299.8) |
| (13) (x2CH4-UV-M7) | 29.6 | 1533 | 3,934,544 | 273.7 (266.9) |
| (14) (x3CH4-UV-M7) | 23.1 | 1502 | 5,468,389 | 267.9 (271.7) |

Run 1 in Table1 reproduces the modern Earth (see also Grenfell et al., 2011 for more details).

Run 2 - features a weak $O_3$ column, but high $N_2O$ and $CH_4$ column values. This result was already shown and discussed e.g. in Segura et al. (2005), Rauer et al. (2011) and Grenfell et al. (2013). Briefly, weak stellar UV leads to weak $O_3$ *formation* via $O_2$ photolysis as part of the Chapman mechanism. $N_2O$ and $CH_4$ abundances are related to the UV environment and the rate of transport upwards from the troposphere (see also Tables 2b and 2c).

Run 3 –The mild $O_3$ increase is linked with stronger Chapman production (see also Table 2a). $N_2O$ and $CH_4$ abundances are reduced compared with run 2 (M7) due to stronger UV-associated sinks. The stratospheric temperature for run 3 is somewhat colder than in run 2 since less $CH_4$ in run 3 leads to less heating ($CH_4$ is a key stratospheric heater in these runs).

Run 4 - column values are more similar to the Earth control than run 2, suggesting that changes in stellar UV can critically affect biosignature abundances. The x10 increased UV in run 4 (x10UV-M7) compared with run 3 (UV-M7) leads to a stronger $O_3$ layer (consistent with more Chapman production) but weaker $N_2O$ and $CH_4$ columns (consistent with more UV-related loss of these species as already discussed).

Run 5 – has some column values which are also rather similar to the Earth control (run 1). Compared with run 4, the increased UV in run 5 has the expected effect upon column $N_2O$ and $CH_4$ i.e. these species are reduced due to increased UV-related sinks (see also Tables 2b, 2c). Regarding $O_3$ run 5 features a stronger $O_3$ layer than run 4 due to stronger Chapman chemistry. A chemical analysis follows in section 3.2 (see also Table 2a).

Run 6 – has a rather strong $O_3$ column although a little weaker than run 5 – this is related e.g. to stronger HOx in run 6 which destroyed some $O_3$ (see also chemical analysis, 3.2).

Run 7 – has somewhat weaker $O_3$ column than run 4. $O_3$ is especially suppressed (e.g. due to stronger $O_3$ photolysis rates) in the lower stratosphere  (see Figure 8 and chemical analysis, section 3.2).

Run 8 – values hardly changed compared with run 3. In both run 3 and run 8 the enhanced Lyman-$\alpha$ radiation was efficiently absorbed in the model's uppermost layers and could not therefore strongly influence the biosignature columns and temperature profiles. In runs 3 and 6, even the uppermost model layer in the chemistry module was already optically thick to incoming Lyman-$\alpha$ radiation. Further investigation showed that $CH_4$ (which is present in large amounts in these runs compared with the Earth control, see Table1) played a major role in this absorption.

Run 9 – lowering surface $CH_4$ emissions by x100 led to an atmospheric column reduction in $CH_4$ but only by a factor of ~8.6 compared with run 2. For a completely passive tracer, the column amount would respond linearly to changes in surface emission. For $CH_4$ however, its weak column response suggests a negative feedback. The following section discusses chemical responses in more detail. Less $CH_4$ in run 9 led to a higher transparence of the atmosphere to incoming radiation which resulted in ~20K surface warming compared with run 2.

Run 10 – the x100 reduced surface $CH_4$ emissions for a planet in a high UV environment compared with run 3 led to a strong lowering - by a factor 70 – in the $CH_4$ atmospheric column. The column response is smaller than the initial change in surface flux, which suggests a negative feedback (see following section on chemical responses).

Run 11 – removing the $N_2O$ surface biomass emissions led to a strong reduction in the $N_2O$ atmospheric column which suggests that $N_2O$ is indeed a good biomarker. In run 11 abiotic reactions which take place in-situ in the atmosphere form $N_2O$ but very slowly compared with biogenic processes.

Run 12 – lowering the $N_2O$ surface biomass emissions by a factor x1000 led to a lowering in the atmospheric column, but only by a factor 89 compared with run 2. The weak column response is discussed further in the chemical responses section (3.2). Absolute $N_2O$ column amounts were still rather high (Table 1) compared with the Earth control scenario with high UV because of the weak UVassociated loss. The surface temperature in run 12 is cool (and the middle atmosphere warm) compared with run 2 (M7) due to a strong increase in $CH_4$ in run 12 (see chemical analysis section 3.2) which radiatively heats the middle atmosphere but leads to a cooling effect in the lower layers since it blocks incoming stellar energy. Note that an initial lowering in $N_2O$ emissions, therefore leads to a cooler surface which could lead to less biological activity (this effect is not included in the present model version). This phenomenon could result in an even stronger decrease in $N_2O$ emissions – a positive feedback.

Runs 13 and 14 – feature somewhat lower $O_3$ columns e.g. because their enhanced $CH_4$ leads to a warmer middle atmospheres via enhanced $CH_4$ heating which suppresses $O_3$ due to an increase in the T-dependent reaction $O_3+O\rightarrow2O_2$.

### 3.2 Photochemical responses to UV radiation and biomass emissions in the mid-stratosphere (~50mb)

In this section we present in detail key photochemical rates (output near 50mb in the low-to-mid stratosphere) affecting $O_3$, $N_2O$ and $CH_4$ in order to understand responses to UV and surface biomass.

### 3.2.1 Responses of $O_3$

The $O_3$ photochemical response to UV radiation is critically sensitive to wavelength. There are three important effects. Firstly, UV ($\lambda<180$-242nm) splits $O_2$ into ground-state oxygen atoms ($O^3P$) which then react with $O_2$ to *form* $O_3$. Secondly, UV ($\lambda<240$-320nm) radiation quickly photolyses $O_3$ into $O_2$ and $O^3P$ which can in turn (quickly) undergo the reverse reaction: $O^3P+O_2+M\rightarrow O_3+M$ (where M denotes any third-body needed to remove excess vibrational energy) to reform $O_3$. Thirdly, depending on wavelength, UV radiation can photolytically destroy so-called reservoir molecules (e.g. nitric acid $HNO_3$, chlorine nitrate, $ClONO_2$) so that they release HOx [$=OH+HO_2$] and NOx [$=NO+NO_2$]. These species can *destroy* (Chapman) $O_3$ via catalytic loss cycles but they may help to form (smog) $O_3$ in low UV environments (see e.g. WMO Report, 1995).

Assuming photochemical steady-state for ground-state ($O^3P$) and excited-state O atoms then, under stratospheric conditions (above ~25km on Earth) yields (see e.g. Brasseur and Solomon, 2005) :

$$[O_3] \sim [(k_{(O3P+O2+M)}/k_{(O3P+O3)})[M][O_2]^2(k_{(O2+hv)}/k_{(O3+hv)})]^{0.5}$$

where k denotes the reaction coefficient for the reaction as indicated by the subscript and square brackets enclosing a species denote abundance. This equation therefore suggests that the abundance of Chapman $O_3$ in steady-state is determined by (i) the fast 3-body rate forming $O_3$, (ii) the $O_2$ abundance, (iii) the bath-gas abundance, M, and (iv) the rate of $O_2$ and $O_3$ photolysis. Regarding the effect of UV radiation upon $O_3$, the equation suggests: (a) [$O^3P$] in the numerator (hence $O_3$ production) is favoured by radiation below ~180nm (which drives: $O_2+hv_{(UVC)}\rightarrow2O^3P$), whereas (b) $k_{(O3+hv)}$ in the equation denominator (which drives: $O_3+hv_{(UVB)}$ photolytic loss) is favoured by UVB radiation at ~300nm. There is also a secondary temperature (T) effect which involves the Chapman $O_3$ removal reaction: ($O^3P+O_3\rightarrow2O_2$). This reaction has a strong, positive, T-dependent rate constant (proportional to exp(-2060/T)). Thus, warm temperatures generally favour $O_3$ removal. In summary, atmospheric $O_3$ can be quickly formed or destroyed depending on wavelength-dependent changes in UV radiation and in the temperature. The issue is also complicated by the fact that $O_3$ is usually itself the major absorber of UV in Earth-like atmospheres and typically dictates the near UV environment in regions below where it forms.

Changing surface biomass emissions of $N_2O$ and $CH_4$ can also influence $O_3$ because e.g. $N_2O$ destruction is a major source of NOx in the stratosphere (leading to $O_3$ loss) and $CH_4$ is an important greenhouse gas which can affect: (1) the T-sensitive Chapman reaction: $O+O_3\rightarrow2O_2$ and (2) HOx (hence $O_3$) since $CH_4$ is the main sink for OH and higher tropospheric T can increase the amount of water vapour in the atmosphere via evaporation.

To investigate these effects, Table 2a shows the concentration and key photochemical rates affecting $O_3$ in the low-to-mid stratosphere (near 50mb). Column 3 shows the rate of $O_2$ photolysis – this process is associated with strong $O_3$ formation since it efficiently produces O atoms which quickly form $O_3$ via the fast reaction: (O+ $O_2+M\rightarrow O_3+M$) as already mentioned. For oxygen photolysis, Table 2a shows only the rate of the reaction channel producing ground-state ($O^3P$) oxygen atoms; the rate of the alternative channel, producing excited-state ($O^1D$) atoms was negligibly low at 50mb. Column 4 shows the rate of the important $O_3$ Chapman sink: $(O+O_3 \rightarrow 2O_2)$, whereas columns 5 and 6 show the rate of $O_3$ loss via photolysis.

Table 2a suggests that runs with enhanced UV (compared with the low UV M7 run) mostly feature enhanced $O_3$ concentrations by (1-2) orders of magnitude. The Chapman production rates (i.e. $O_2+hv$ shown in column 3, Table 2a) is strongly stimulated for run 6 with x1000 UV. Results therefore suggest that an important effect of enhanced UV is to favour atmospheric $O_3$; there occurs however a minor, opposing effect, whereby $HOx$ (which destroys ozone) increased (see Table 2c) at enhanced UV. This effect was important e.g. for run 6, Table 2c). The effect of increasing UV in run 7 (see also Figure 2) leads to a reduction in $O_3$ by a factor x2 to x3 e.g. due to enhanced $O_3$ photolysis (see also $O_3$ profiles, Figure 8). The effect of increasing Lyman-$\alpha$ in run 8 has only a small effect (mostly <1%) compared with run 3, as mentioned.

The effect of reducing surface biomass emissions in Table 2a upon $O_3$ generally leads to rather small $O_3$ values.. A rather large reduction in $O_3$ occurs in run 12 i.e. with reduced $N_2O$ compared with run 2 (M7). In these low-UV runs, the smog mechanism (Haagen-Smit, 1952) is an important $O_3$ source (Grenfell et al., 2013) since UV levels are not sufficient to drive efficiently the Chapman mechanism. The especially low $O_3$ abundance in run 12 therefore arises because reducing $N_2O$ leads to a reduction in NOx compared with run 2 (M7) (from ~49 down to 13 ppb near 50mb) since NOx is formed from $N_2O$ decomposition. This NOx reduction slows the $O_3$ smog mechanism. For runs 13 and 14, increasing $CH_4$ leads to a warmer middle atmosphere which stimulates the reaction $O_3+O$ hence reduces $O_3$, as discussed.

Table 2a: Chemical abundance [ppb] and rates (ppb/s) of key photochemical sources and sinks affecting $O_3$. Data are output for the model (variable) gridpoint closest to 50mb as in Table 1. $O^3P$=ground-state, $O^1D$=excited state oxygen atoms.

| Run | [$O_3$] | ($O_2+hv \rightarrow 2O^3P$) | ($O_3+O^3P \rightarrow 2O_2$) | ($O_3+hv \rightarrow O^3P+O_2$) | ($O_3+hv \rightarrow O^1D+O_2$) |
|---|---|---|---|---|---|
| (1) Earth control | 2.5E+03 | 3.2E-05 | 2.3E-06 | 6.7E-01 | 4.0E-02 |
| (2) M7 | 4.9E+02 | 1.3E-06 | 4.1E-07 | 1.4E-02 | 3.1E-04 |
| (3) UV-M7 | 8.0E+02 | 4.4E-05 | 4.9E-07 | 2.3E-02 | 5.1E-04 |
| (4) x10-UV-M7 | 2.7E+03 | 1.0E-06 | 9.5E-07 | 7.5E-02 | 3.2E-04 |
| (5) x100-UV-M7 | 4.2E+03 | 2.7E-07 | 1.4E-06 | 1.2E-01 | 9.1E-04 |
| (6) x1000-UV-M7 | 1.0E+04 | 2.8E-04 | 3.0E-05 | 2.9E-01 | 7.3E-02 |
| (7) x10-UV-M7x5-slope | 9.8E+02 | 1.9E-06 | 1.1E-07 | 2.8E-02 | 1.1E-03 |
| (8) x100-Ly-$\alpha$-UV-M7 | 8.0E+02 | 4.4E-05 | 4.9E-07 | 2.3E-02 | 5.1E-04 |
| (9) LOW-CH4-M7 | 3.2E+02 | 4.5E-07 | 2.6E-08 | 9.3E-03 | 1.3E-04 |
| (10) LOW-CH4-UV-M7 | 7.6E+02 | 6.3E-07 | 2.9E-08 | 2.2E-02 | 4.8E-05 |
| (11) ABIO-N2O-UV-M7 | 7.6E+02 | 4.2E-05 | 4.7E-07 | 2.2E-02 | 4.4E-04 |
| (12) LOW-N2O-M7 | 1.3E+01 | 4.5E-06 | 3.4E-09 | 3.8E-04 | 3.2E-05 |
| (13) UV-M7-x2CH4 | 7.30E+02 | 1.10E-04 | 1.00E-06 | 2.10E-02 | 1.40E-03 |
| (14) UV-M7-x3CH4 | 5.90E+02 | 1.40E-04 | 9.20E-07 | 1.70E-02 | 1.90E-03 |

### 3.2.2 Responses of $N_2O$

UV radiation destroys $N_2O$ (typically on Earth in the mid-to-upper stratosphere), either (i) via direct photolysis in the wavelength region: ($195nm < \lambda < 205nm$), or (ii) via reaction of $N_2O$ with $O^1D$ (which is formed from UV-photolysis of $O_3$). There exist two ($N_2O+O^1D$) reaction product channels forming (i) 2NO (which can go on to destroy catalytically stratospheric $O_3$) and (ii) $N_2+O_2$. Regarding the effect of biomass emissions on $N_2O$ abundance, clearly a good test of a biomarker is that its modelled atmospheric abundance must respond strongly to a removal in its surface biomass source (this may not occur e.g. if non-linear atmospheric feedbacks are significant).

To investigate these effects, Table 2b shows concentration and key photochemical rates affecting $N_2O$, output again near 50mb. Results imply larger $N_2O$ abundances for the low-UV (e.g. M7) scenarios, since UV favours $N_2O$ removal. Run 5 features a similar $N_2O$ abundance as the Earth control, but run 6, with its enhanced UV, features significantly less $N_2O$ than the Earth. The reaction rates imply that photolysis is mostly a faster $N_2O$ sink than the reaction with $O^1D$. Increasing the incoming TOA Lyman-$\alpha$ fluxes by x100 (run 8) has a negligible effect compared with run 3 (i.e. a similar result as for the $O_3$ case in Table 2a, and for similar reasons).

Regarding the effect of surface biomass, changing the $CH_4$ biomass in Table 2b has only a small effect on $N_2O$. Switching off the surface $N_2O$ biomass emissions (run 11) (i.e. so that only slow, abiotic atmospheric sources remain e.g. via the reaction: $N_2+O^1D+M\rightarrow N_2O+M$) leads to a strong reduction in the abundance of $N_2O$ by more than 5 orders of magnitude compared with run 3 (UV-M7). $N_2O$ is therefore a "good biomarker" in this respect. Finally, decreasing the surface $N_2O$ biomass by a factor x1000 in run 12 compared with to run 2 (M7) leads to a decrease in the $N_2O$ abundance - but "only" by a factor 89. This non-linearity suggests processes occurring which oppose the surface biomass lowering of $N_2O$ in run 12. One important effect was the change in abundance of electronically excited oxygen atoms ($O^1D$) (a sink for $N_2O$) which decreased by a factor 48 in run12 compared with run 2 near 50mb. This arose since less $O_3$ (as discussed) in run 12 led to weaker $O^1D$ production via $O_3$ photolysis.

Table 2b: Chemical abundances [ppb] and rates (ppb/s) of key photochemical processes affecting $N_2O$. Data are output for the model (variable) gridpoint closest to 50mb which corresponds to the low to mid-stratosphere for Earth.

| Run | [$N_2O$] | ($N_2O+hv\rightarrow N_2+O$) | ($N_2O+O^1D\rightarrow 2NO$) | ($N_2O+O^1D\rightarrow N_2+O_2$) |
|---|---|---|---|---|
| (1) Earth control | 2.30E+02 | 4.40E-08 | 1.90E-08 | 1.40E-08 |
| (2) M7 | 3.20E+05 | 5.40E-07 | 2.10E-07 | 1.60E-07 |
| (3) UV-M7 | 1.80E+03 | 3.30E-07 | 2.00E-09 | 1.50E-09 |
| (4) x10-UV-M7 | 6.20E+02 | 4.60E-09 | 4.30E-10 | 3.10E-10 |
| (5) x100-UV-M7 | 3.30E+02 | 6.80E-10 | 6.30E-10 | 4.80E-10 |
| (6) x1000-UV-M7 | 1.20E+02 | 2.10E-07 | 1.90E-08 | 1.40E-08 |
| (7) x10-UV-M7-x5Slope | 5.60E+02 | 8.00E-09 | 1.40E-09 | 1.00E-09 |
| (8) x100-Ly-α-UV-M7 | 1.80E+03 | 3.30E-07 | 2.00E-09 | 1.50E-09 |
| (9) LOW-CH4-M7 | 3.20E+05 | 2.30E-07 | 9.10E-08 | 6.60E-08 |
| (10) LOW-CH4-UV-M7 | 2.30E+02 | 8.90E-09 | 2.20E-10 | 1.60E-10 |
| (11) ABIO-N2O-UV-M7 | 1.50E-02 | 1.10E-12 | 5.60E-15 | 4.10E-15 |
| (12) LOW-N2O-M7 | 3.60E+03 | 1.50E-08 | 2.40E-10 | 1.70E-10 |
| (13) UV-M7-x2CH4 | 1.70E+03 | 5.90E-07 | 5.00E-09 | 3.70E-09 |
| (14) UV-M7-x3CH4 | 1.70E+03 | 6.60E-07 | 6.80E-09 | 5.00E-09 |

### 3.2.3 Responses of $CH_4$

UV radiation is usually associated with $CH_4$ destruction in Earth's atmosphere. This is because $CH_4$ is removed mainly by the hydroxyl (OH) radical via the reaction: $CH_4 +OH\rightarrow CH_3+ H_2O$ (in fact, this is the first in a series of reactions whereby $CH_4$ may be completely oxidized into $CO_2$). OH is usually enhanced by UV since $O^1D$ (which comes mainly from UV-photolysis of $O_3$) features in the key OH source reaction: $H_2O+ O^1D \rightarrow 2OH$. The abundance of OH is also coupled with nitrogen chemistry via the fast reaction: $NO+HO_2 \rightarrow NO_2+OH$. In summary, enhanced UV conditions (on Earth) normally lead to increased OH therefore reduced $CH_4$.

Table 2c shows the concentration and key photochemical rates affecting the $CH_4$ abundance, again output near 50mb. On Earth, enhanced UV usually leads to more OH since this species is photolytically produced. This is indeed the case in going from runs 4 to 6 (Table 2c) but is not the case from runs 2 to 4, related to a shift in OH related to changes in NOx (see Grenfell al., 2013 for further details). Finally, Table 2c suggests that OH is mostly the major sink for $CH_4$ (e.g. Cl plays a relatively small role).

The effect of changing biomass in Table 2c is shown in runs 9-14. In run 9, lowering surface biomass $CH_4$ emissions by a factor of x100 leads to a lowering in $CH_4$ abundance at 50mb, but only by a factor 70 compared with run 2. This suggests a *negative* atmospheric feedback i.e. opposing the effect of lowering by x100 the surface emissions and resulting in "only" a factor of 70 lowering in abundance. On Earth, however, the situation is different. Here, an increase in atmospheric $CH_4$ (due to increasing man-made activities) is usually discussed to lead to a decrease in OH, since $CH_4$ and OH on Earth are normally each other's main atmospheric

sink – this is a *positive* feedback (Prather, 1996). The behavior of run 9 therefore suggests fundamental feedback differences between $CH_4$ responses compared with the Earth. In run 9, the lowered $CH_4$ leads to a cooler stratosphere by about 35K at 50mb (Table 1). This slows the rate constant of the reaction: $(CH_4+OH)$ which has a T-dependence proportional to (-1775/T), but not for other competing sinks such as the reaction $(CH_4+O^1D)$ whose rate constant does not have a T-dependence. Furthermore, less $CH_4$ in run 9 compared with run 2 means less $H_2O$ in the stratosphere (by a factor 2-3) due to weaker $CH_4$ oxidation which is an important $H_2O$ source there. This favours a weakening in the OH source reaction: $(H_2O+O^1D)$. The result of these two effects is that OH in run 9 (unlike on Earth) becomes weak and is indeed no longer the dominant photochemical sink of $CH_4$ as shown in Table 2c. In summary, reduced $CH_4$ in run 9 leads to a slowing in its removal rate via OH which opposes the original $CH_4$ reduction – a negative feedback.

In run 10, reducing the surface $CH_4$ biomass by x100 in run 3 (UV-M7), results in the $CH_4$ abundance at 50mb decreasing by a factor 70 – this is a negative feedback for similar reasons as discussed for run 9.. Run 11, compared with run 3 has only a very small effect on $CH_4$ so is not discussed further. Run 12, in which the surface $N_2O$ biomass source was reduced by x1000 compared with run 2 (M7) features a large $CH_4$ abundance associated with lowered OH (its main sink) which arose e.g. due to a weakening in the OH chemical sources (Table 2c). In runs 13 and 14, doubling and tripling the $CH_4$ surface biomass sources respectively led to an (approximate) doubling and tripling in the $CH_4$ abundance – this suggested that any feedbacks operating were weak.

Table 2c: Chemical abundance [ppb] and rates (ppb/s) of key photochemical processes affecting $CH_4$. Data are output for the model (variable) gridpoint closest to 50mb which corresponds to the low to mid-stratosphere for Earth. Pressure values vary from 48.2mb (Earth control run) up to 49.9mb (LOW-CH4-UV run). Note that the reaction $CH_4+O^1D$ has two channels forming: (1) $CH_3+OH$ and (2) $HCHO+H_2$ respectively.

| Run | [CH4] | (H2O+O1D) | (HO2) | (OH) | (O1D) | (NO+HO2) | (CH4+OH) | (CH4+O1D) Channel 1 | (CH4+O1D) Channel 2 | (CH4+Cl) |
|---|---|---|---|---|---|---|---|---|---|---|
| (1) Earth control | 1.3E+03 | 1.8E-06 | 2.3E-03 | 4.3E-04 | 7.9E-10 | 2.8E-05 | 5.3E-07 | 2.2E-07 | 2.2E-08 | 5.1E-08 |
| (2) M7 | 5.7E+06 | 1.7E-06 | 1.8E-01 | 6.8E-06 | 2.2E-11 | 1.3E-04 | 6.0E-05 | 8.0E-06 | 8.0E-07 | 2.2E-07 |
| (3) UV-M7 | 2.6E+06 | 4.7E-06 | 1.1E-01 | 5.2E-06 | 2.7E-11 | 6.0E-06 | 2.1E-05 | 6.1E-06 | 6.1E-07 | 1.4E-07 |
| (4) x10-UV-M7 | 2.3E+05 | 5.0E-07 | 6.6E-03 | 2.3E-06 | 7.9E-12 | 2.5E-08 | 9.4E-07 | 3.3E-07 | 3.3E-08 | 7.0E-08 |
| (5) x100-UV-M7 | 5.2E+04 | 7.9E-07 | 5.1E-03 | 6.8E-06 | 1.9E-11 | 8.8E-10 | 5.5E-07 | 2.1E-07 | 2.1E-08 | 1.6E-07 |
| (6) x1000-UV-M7 | 1.7E+04 | 4.7E-05 | 4.7E-02 | 1.2E-03 | 3.2E-09 | 1.2E-07 | 2.0E-05 | 5.4E-06 | 5.4E-07 | 6.6E-07 |
| (7) x10-UV-M7-x5Slope | 1.7E+06 | 1.2E-06 | 2.1E-02 | 6.4E-06 | 2.3E-11 | 3.6E-07 | 2.1E-06 | 8.4E-07 | 8.4E-08 | 1.5E-07 |
| (8) x100-Ly-α-UV-M7 | 2.6E+06 | 4.7E-06 | 1.1E-01 | 5.2E-06 | 2.7E-11 | 6.0E-06 | 2.1E-05 | 6.1E-06 | 6.1E-07 | 1.4E-07 |
| (9) LOW-CH4-M7 | 6.2E+05 | 2.8E-07 | 4.1E-02 | 2.8E-09 | 4.5E-12 | 2.9E-05 | 2.7E-09 | 3.7E-07 | 3.7E-08 | 1.3E-08 |
| (10) LOW-CH4-UV-M7 | 3.7E+04 | 1.5E-08 | 6.6E-03 | 1.5E-07 | 8.1E-13 | 2.1E-08 | 9.1E-09 | 7.7E-09 | 7.7E-10 | 1.8E-08 |
| (11) ABIO-N2O-UV-M7 | 2.9E+06 | 4.2E-06 | 1.1E-02 | 4.4E-06 | 2.4E-11 | 5.2E-07 | 2.0E-05 | 5.9E-06 | 5.9E-07 | 1.3E-07 |
| (12) LOW-N2O-M7 | 2.7E+07 | 3.5E-08 | 8.0E-01 | 2.6E-06 | 5.5E-12 | 2.7E-06 | 1.1E-04 | 3.7E-06 | 3.7E-07 | 2.5E-07 |
| (13) UV-M7-x2CH4 | 4.9E+06 | 8.1E-06 | 2.7E-01 | 1.1E-05 | 1.0E-10 | 8.1E-06 | 8.6E-05 | 3.1E-05 | 3.1E-06 | 2.8E-07 |
| (14) UV-M7-x3CH4 | 6.9E+06 | 1.0E-05 | 4.1E-01 | 1.4E-05 | 1.5E-10 | 1.6E-05 | 1.7E-04 | 5.9E-05 | 5.9E-06 | 3.4E-07 |

## 3.3 Atmospheric Profile Responses

### 3.3.1 Effect of stellar UV and Lyman-α emission

Figure 3 shows upward and downward shortwave (SW) radiative fluxes from the climate module (considered for λ>238nm see 2.1) for runs (1-6). The downward TOA SW flux for different runs is progressively increased in going from the high UV runs (e.g. x1000UV-M7) to the low UV runs (e.g. M7) due to the scaling procedure discussed in section 2.3. The upward fluxes in Figure 2 are reduced for the M-dwarf scenarios compared with the Earth control mostly due to weaker Rayleigh scattering.. In the lower atmosphere in Figure 2 the downward shortwave fluxes are reduced due to atmospheric absorption especially for the M7 run which features strong $CH_4$ absorption.

Figures 4 and 5 show UV radiation fluxes in the wavelength range UVB ($280<\lambda<315$nm) and UVC ($176<\lambda<280$nm) respectively. Figure 4 shows that with increasing TOA UVB, radiation penetrates downwards through the atmosphere until a point is reached in the stratosphere where e.g. the UV flux and [$O_2$] conditions lead to an $O_3$ layer forming (see also Figure 8) – below this level the UVB is then efficiently absorbed and falls off quickly. One sees that with increasing TOA UVB, the radiation can penetrate to successively lower altitudes before $O_3$ absorption occurs, which leads to a sudden change in the UVB gradient profile. For the M7 and UV-M7 cases, however, the incoming radiation is too weak to form a strong $O_3$ layer and these cases show more isoprofile behaviour.. Figure 5 (UVC) shows some similarities with Figure 4 i.e. once an $O_3$ layer forms, the incoming UVC (which is critical to drive the Chapman mechanism hence to form $O_3$) is absorbed and there occurs a strong drop-off in UVC radiation at lower altitudes.

Figure6 shows calculated heating rate profiles. Strong heating rates are calculated in the middle atmosphere (i) in the M7 run due to $CH_4$ heating and (ii) in the x1000-UV-M7 run caused by enhanced UV radiation and ozone amounts.

Figure 7 shows temperature profiles. The Earth control run and the x1000-UV-M7 run both have strong temperature inversions due to (i) strong $O_3$ layers and (ii) sufficient UV stellar irradiation at wavelengths where $O_3$ absorbs. Remaining runs have weak temperature inversions and cool stratospheres for similar reasons except for runs with M7 stellar spectra, which have significant $CH_4$ heating in the stratosphere.

Figures (8-10) show the $O_3$, $N_2O$ and $CH_4$ atmospheric profiles respectively for all runs. In Figure 8, the M7 run has a weak $O_3$ abundance since the low UV disfavours Chapman production. On increasing the UV from UV-M7 through x10UV-M7 and up to x1000-UV-M7, $O_3$ layers are formed with peaks which are somewhat narrower in the vertical and which form increasingly lower in altitude as UV increases. In run 7, the effect of changing UV is strongest in the low to mid stratosphere (compare green continuous and green dashed lines in Figure 8).

Figure 9 suggests that $N_2O$ is removed faster as UV is increased from run to run. This is especially evident on the uppermost levels where enhancing UV leads to a steeper decline in abundance gradients.

Figure 10 implies that $CH_4$ is increased by more than three orders of magnitude for the M7 run compared with the Earth control (as also shown in Rauer et al., 2011). Gradually increasing UV from x10UV-M7 up to x1000-UV-M7 leads to decreasing $CH_4$ since UV generally stimulates $CH_4$ sinks e.g. OH. The x1000UV-M7 run therefore features low $CH_4$ in the middle atmosphere and above. Finally, the effect of varying Lyman-$\alpha$ in run 8 was small (<1% change in abundances and temperature) compared with run 3 since this radiation was absorbed in the uppermost atmospheric layers as discussed.

### 3.3.2. Effect of Surface Biomass

$CH_4$ and $N_2O$ abundances in runs (9-14) were mostly isoprofile in the troposphere and they changed for the most part in proportion to the column responses already discussed in the previous section and shown in Tables 2a-2c. In Figure 8, lowering $N_2O$ emissions in the M7 run by a factor of x1000 (run 12, pink line) led to a strong reduction in $O_3$ – this run featured the lowest $O_3$ of all scenarios. Since $O_3$ in low UV runs is mostly smog-produced, decreasing $N_2O$ (hence NOx which is formed form the photolysis of $N_2O$) decreases the formation of $O_3$ via a slowing in the NOx-catalysed $O_3$ smog mechanism. Reducing $N_2O$ in the UV-M7 run (run 11, orange line), however, yielded in contrast only a small effect, since the smog mechanism was now less important. Reducing surface $CH_4$ emissions led only to modest changes in $O_3$ (compare e.g. the grey line (run 9 LOW-CH4-M7) with the red (M7) lines in Figure 8) but see also the following discussion regarding the effect of biomass on spectral responses.

## 3.4 Spectral Responses

### 3.4.1 Effect of stellar UV and Lyman-$\alpha$ input

We compare theoretical emission spectra for the low UV (M7) run with the high UV (x10UV-M7, x1000UV-M7) runs over the wavelength range (2-20μm). Results from the x100Lyman-$\alpha$-UV-M7 run (not shown) were similar to run 3. Figure 11 (upper panel: 2-5μm, lower panel: 5-20μm) shows spectra for the M7 run, the x10UV-M7 run and the x1000UV-M7 run. The x10UV-M7 and x1000UV-M7 runs, with their enhanced UV irradiation have many more spectral features apparent. We will now discuss two key bands in more detail, namely for $CO_2$ and $O_3$.

Figure 12 is as for Figure 11 but focuses on the $CO_2$ 15μm band for the M7, x10UV-M7 and x1000UV-M7 runs. Many more spectral features are evident for the enhanced UV runs compared with the M7 run which is partly related to the stronger temperature gradients (Figure 7) for the high UV runs. The x1000UV-M7 run features a strong inversion at the $CO_2$ 15μm band centre due to strong stratospheric heating.

Figure 13 is as for Figure 12 but focuses on the $O_3$ 9.6μm band. For the M7 run, the incoming stellar UV was weak, which disfavoured Chapman $O_3$ production. For the x1000UV-M7 run, however, the strong $O_3$ layer was associated with enhanced stratospheric heating which weakened the tropospheric-stratospheric temperature gradient hence weakened the $O_3$ spectral features. A similar effect was noted e.g. in Selsis (2000). For Earth, the $O_3$ band originates at pressures from 0.01 to 0.1 bar (Figure 8) which correspond to temperatures of about 215 K (Figure 7). The band centre is optically thick and radiates at ~215K whereas in the optically thin band wings the radiation from the surface (T~288K) can penetrate the atmosphere. This implies a temperature difference of ~73K between the band centre and its wings. For the x1000UV-M7 run, however, where the band centre originates at p~0.01 bar, an inspection of the temperature profile between surface and stratosphere (Figure 7) shows a much weaker temperature difference due to stronger stratospheric $O_3$ heating in this run, which leads to a much weakened band. The central spectral feature visible in Figure 13 for the x10UV run arises due to the shape of the $O_3$ IR vibrational-rotational transitions. There occur asymmetric-top P- and R-absorption branches (at ~9.7 and ~9.5μm respectively) with a weak central Q branch absorption at ~9.6μm which in the (emission) spectra, corresponds to a maximum.

### 3.4.2 Effect of Surface biomass

Figure 14 compares planet/star flux contrast ratios of the theoretical emission spectra of the M7 run (the same run was also shown in Rauer et al., 2011, as indicated) with the LOW-CH4-M7 run. Results suggest that the $O_3$ band for LOW-CH4-M7 (with its x100 lowered $CH_4$ surface emissions) is still rather weak (although a little stronger than in Rauer et al. 2011) since the $O_3$ column is weak. However, a strong $N_2O$ band is now seen very clearly in Figure 14 which was not seen in the Rauer et al. (2011) M7 run despite the high $N_2O$ abundance since stratospheric $CH_4$ heating weakened this band in their work. Supporting this result, the LOW-CH4-M7 run featured colder temperatures by 20-30K throughout the middle atmosphere compared with the M7 run, which enabled the $N_2O$ feature to appear.

### 4. Discussion and Conclusions

Spectral detection of biosignatures is very challenging, even for favoured targets such as Earth-like planets orbiting in the HZ of cool M-dwarf stars. Knowledge of the UV emission spectra of such stars is highly desirable, since this determines e.g. whether $O_3$ (which critically influences the planetary temperature profile, UV environment and spectral bands) is chemically *created* (since Chapman formation of $O_3$ via $O_2$ photolysis increases in the Herzberg continuum) or *destroyed* (e.g. directly via $O_3$ photolysis or via photolytic release of HOx and NOx). The UVemission of the star also strongly affects planetary $CH_4$, and $N_2O$ photochemistry and climate. $CH_4$ biomass surface emissions above ~1% those of Earth may lead to weak atmospheric vertical temperature gradients which could dampen spectral features of biosignatures such as $N_2O$. The key conclusions of this paper are:

-spectral atmospheric biosignatures responded very sensitively to UV emissions of the central M-dwarf star.

-$O_3$ is favoured by high UV(C) radiation which stimulates its formation, whereas high UV(B) levels can also lead to $O_3$ loss via photolysis, to strong in-situ stratospheric heating and can favour stronger chemical sinks due to release of hydrogen oxides. The latter three effects disfavor $O_3$ spectral detection. Thus stellar UV fluxes must be known to interpret correctly measured spectra of planets orbiting M-dwarf stars.

-reducing $CH_4$ surface emissions for Earth-like planets orbiting cool M-dwarf stars leads to stratospheric cooling which stimulates vertical temperature gradients and may lead to strong stimulation of $N_2O$ spectral bands (assuming an Earth-like biomass).

-increasing TOA incoming Lyman-$\alpha$ fluxes by a factor of x100 had little effect on biosignatures since this radiation is efficiently absorbed in high atmospheric layers.

**Acknowledgements**
This research has been partly supported by the Helmholtz Gemeinschaft (HGF) through the HGF research alliance "Planetary Evolution and Life. We thank J. Stock for useful discussion.

**Figures**

Figure 1: Planetary Top-of-Atmosphere (TOA) incoming high resolution stellar spectral flux ($Wm^{-2}nm^{-1}$).

Figure 2: As for Figure 1 but showing TOA stellar spectral flux (photons $cm^{-2}$ $s^{-1}$) for the x10-UV-M7 run and the same run but with UV (300-350nm) further increased by x5 to test the effect of changing the slope of the flux in this critical region. Data is shown interpolated to the (108) wavelength intervals employed in the chemistry module.

Figure 3: Upwelling (up) and downwards (dn) shortwave flux ($Wm^{-2}$) profiles

Figure 4: Planetary $UVB_{280<\lambda<315nm}$ radiation (W $m^{-2}$) profile.

Figure 5: Planetary $UVC_{176<\lambda<280nm}$ radiation (W $m^{-2}$) profile.

Figure 6: Heating rates (K $day^{-1}$) profiles.

Figure 7: Atmospheric Temperature (K) profiles.

Figure 8: Logarithm ($log_{10}$) of ozone abundance profiles in parts per million (ppm). Note that the ???blue and yellow lines overlie.

Figure 9: Logarithm ($log_{10}$) of nitrous oxide abundance (ppm) profiles. Note that the grey and red lines, overlie.

Figure 10: Logarithm ($log_{10}$) of methane abundance (ppm) profiles. Note that the blue and orange lines overlie.

Figure 11: Theoretical emission spectral radiance ($Wm^{-2}sr^{-1}\mu m^{-1}$). Upper panel: (2-5)$\mu$m, lower panel: (5-20)$\mu$m. Wavelength resolution R=$\lambda/\delta\lambda$=100.0.

Figure 12: Theoretical emission spectral radiance ($Wm^{-2}sr^{-1}\mu m^{-1}$) centred at 15$\mu$m (carbon dioxide fundamental). Wavelength resolution R=$\lambda/\delta\lambda$=100.0.

Figure 13: Theoretical emission spectral radiance ($Wm^{-2}sr^{-1}\mu m^{-1}$) centred at 9.6$\mu$m (ozone fundamental). Wavelength resolution R=$\lambda/\delta\lambda$=100.0.

Figure 14: Theoretical planet/star contrast ratios for the ozone and nitrous oxide band for reduced $CH_4$ emissions.

Figure 1

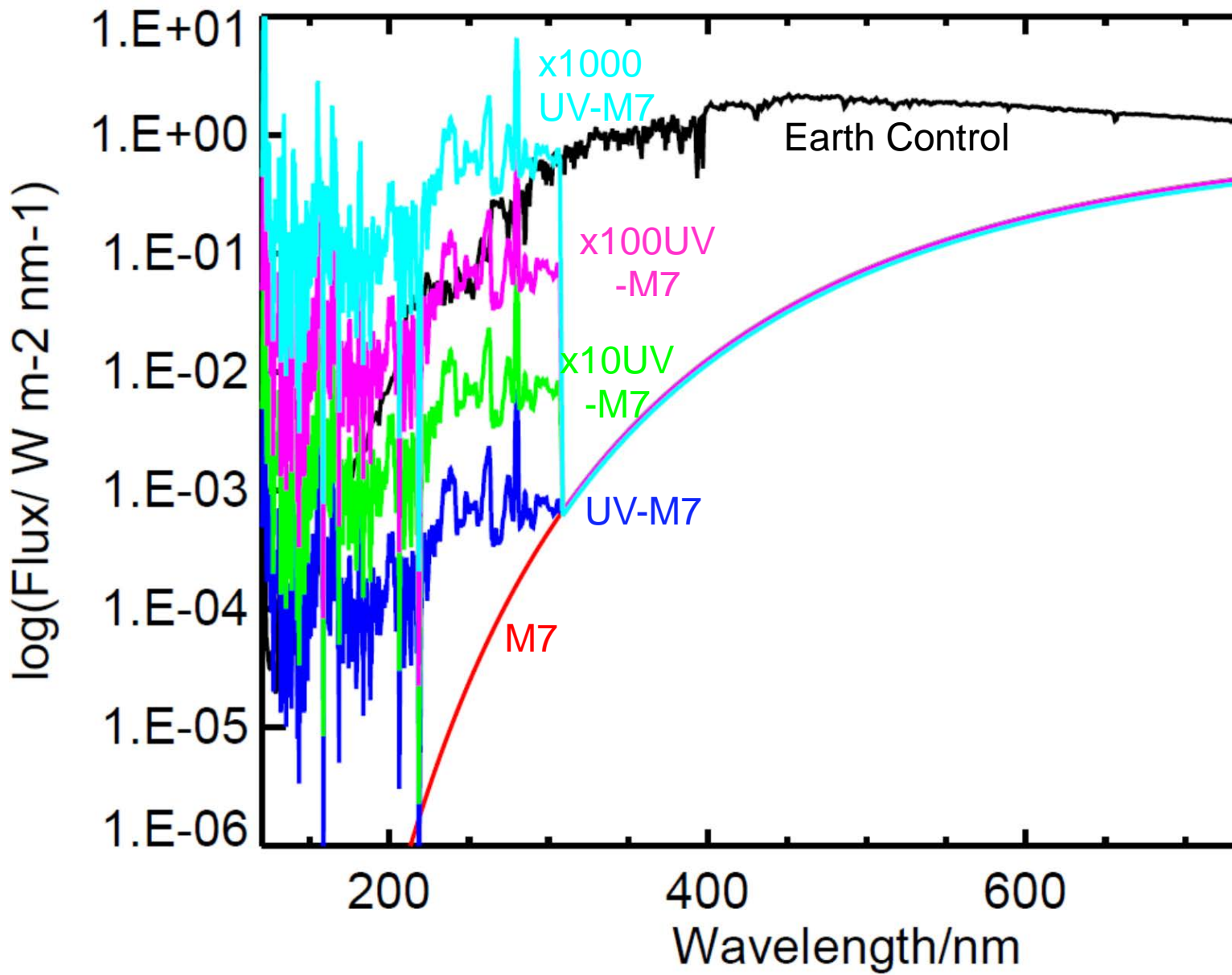

Figure 2

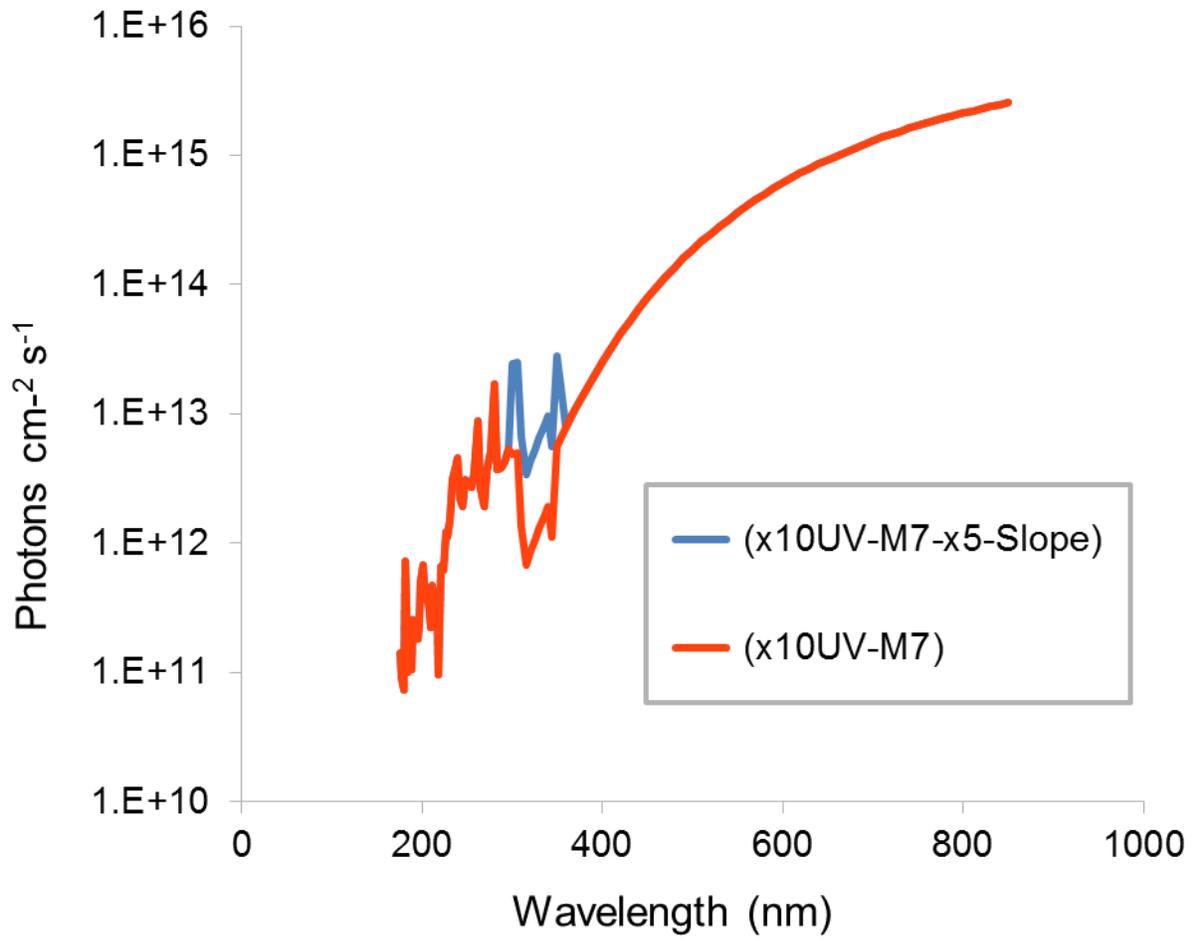

# Figure 3

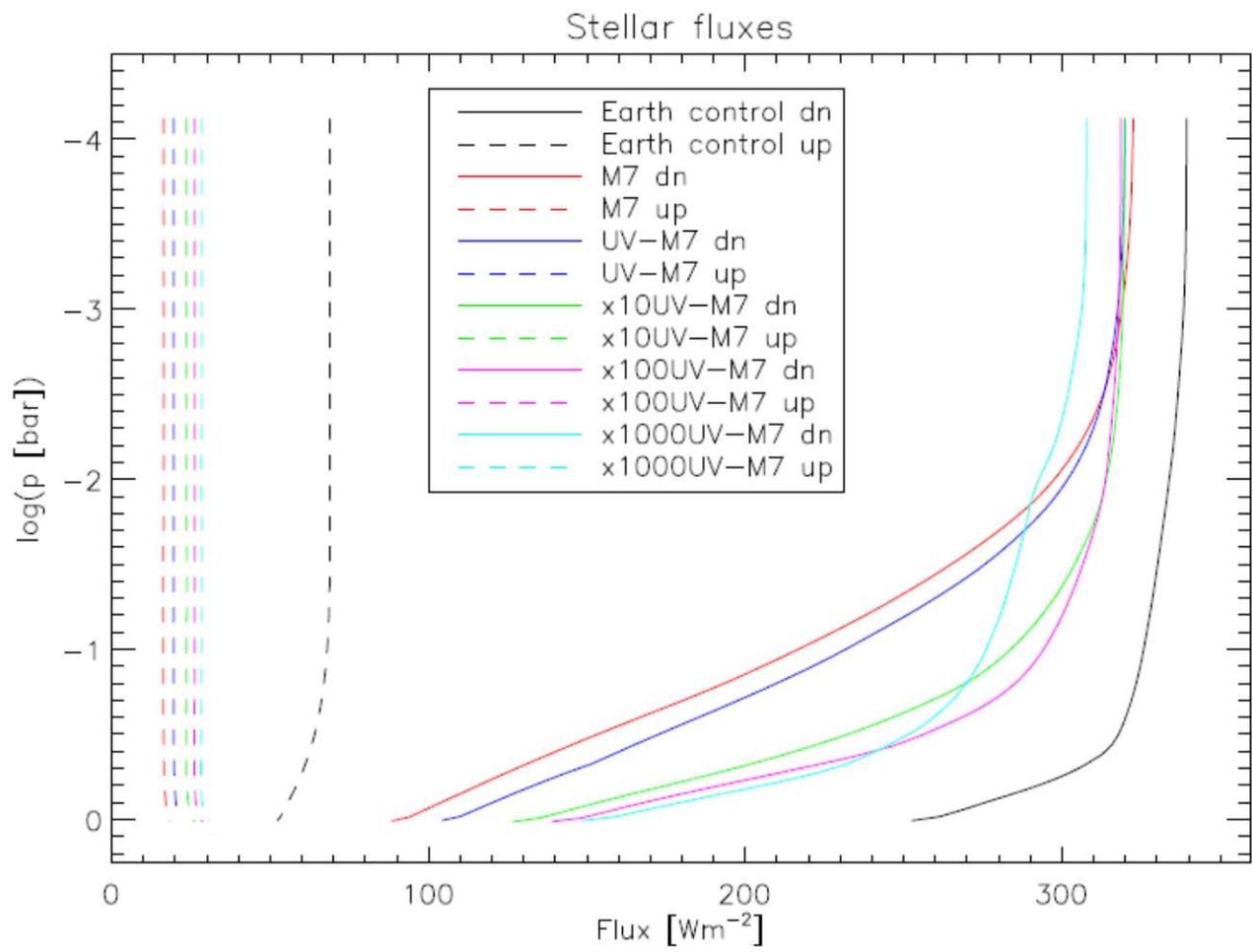

Figure 4

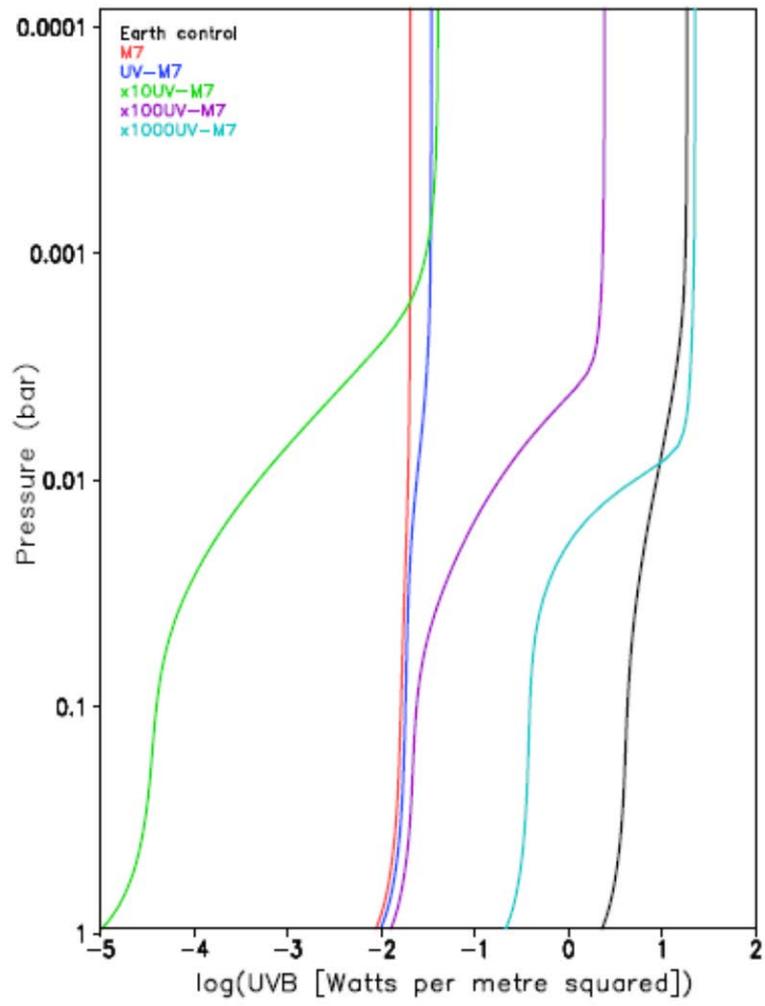

# Figure 5

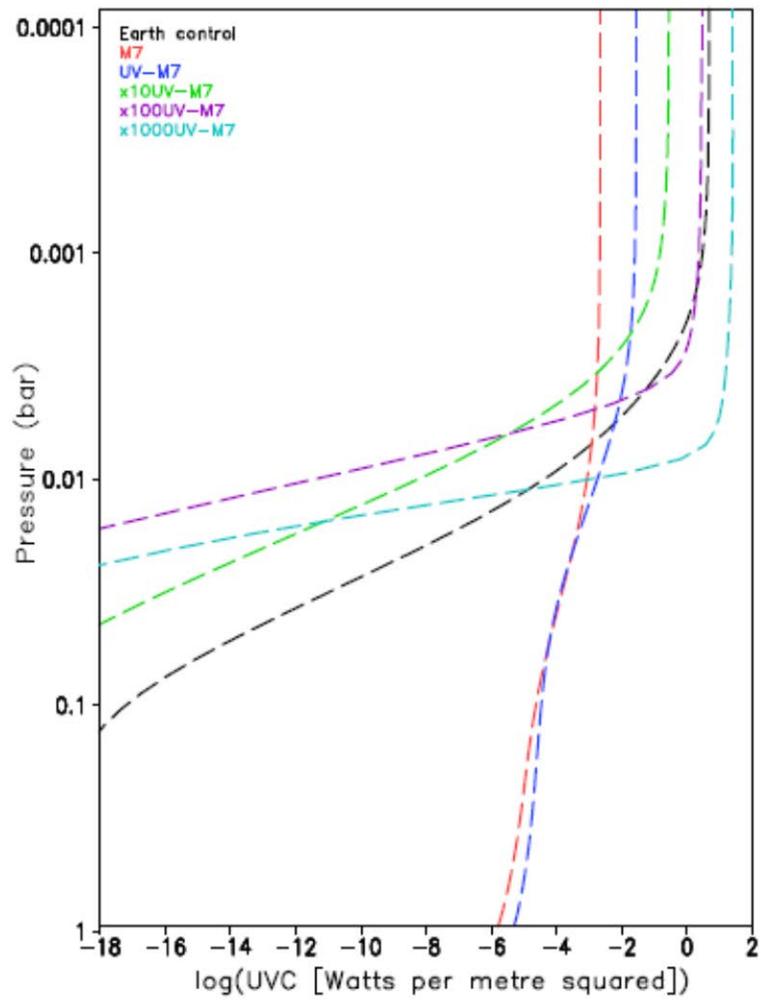

Figure 6

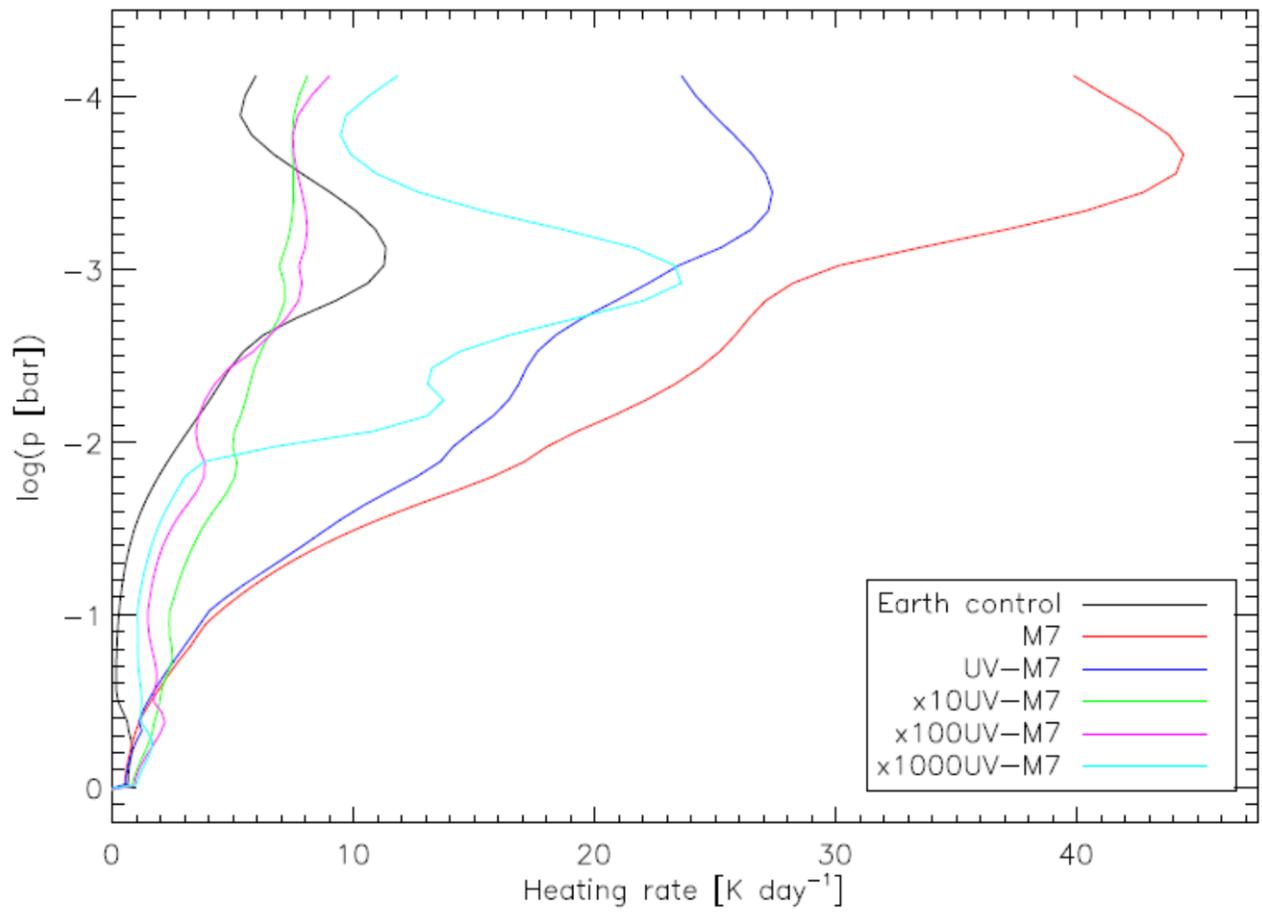

# Figure 7

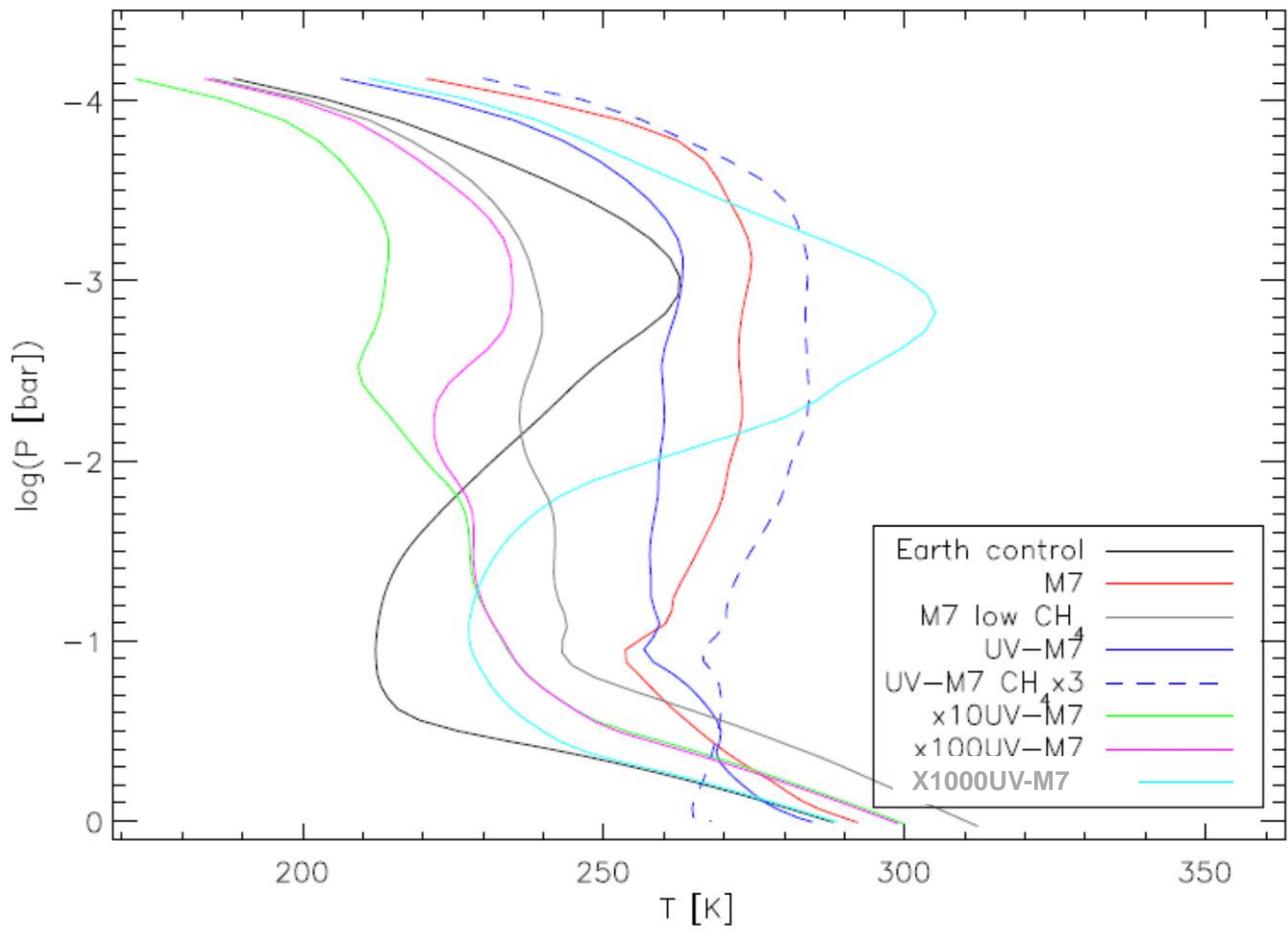

Figure 8

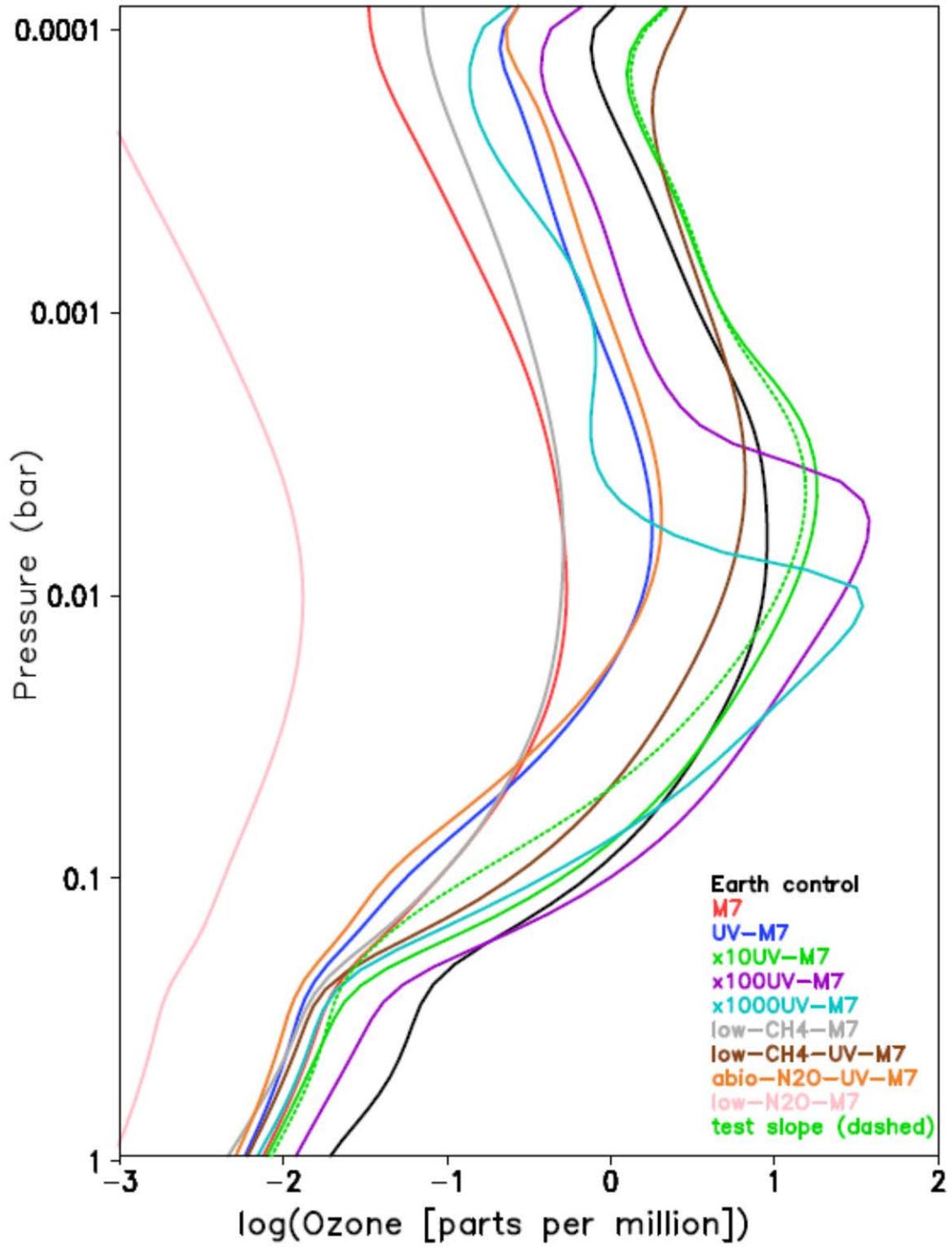

Figure 9

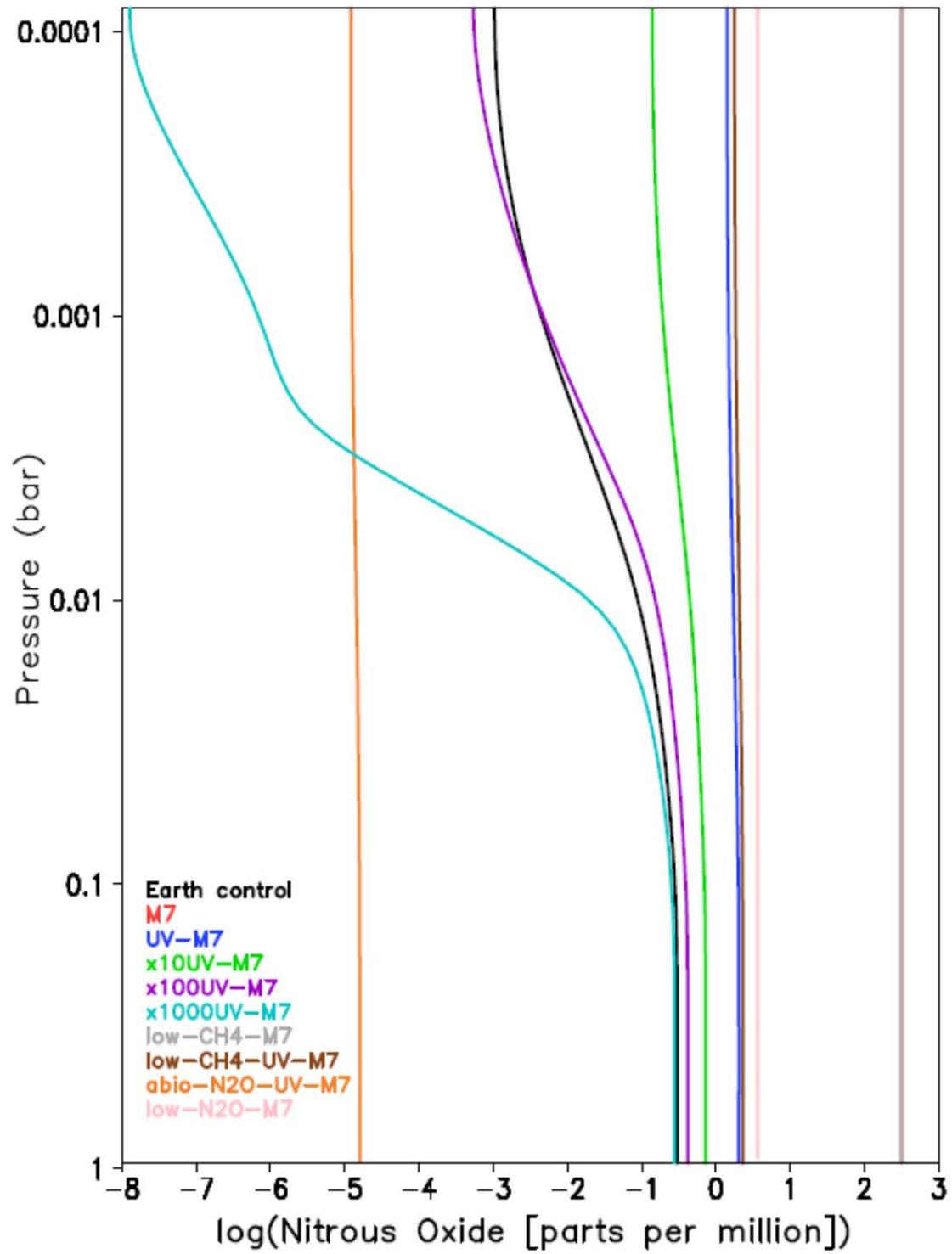

Figure 10

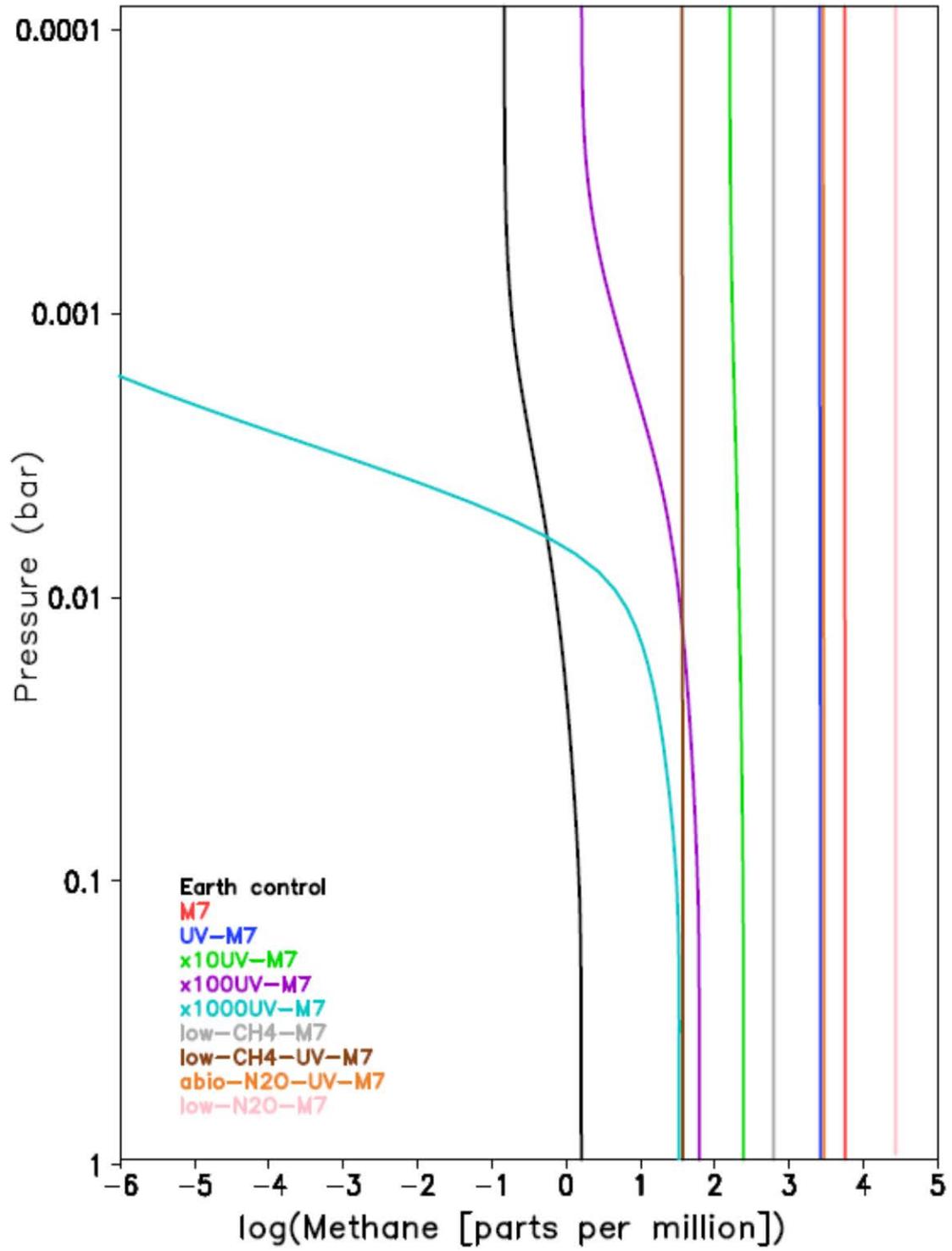

# Figure 11

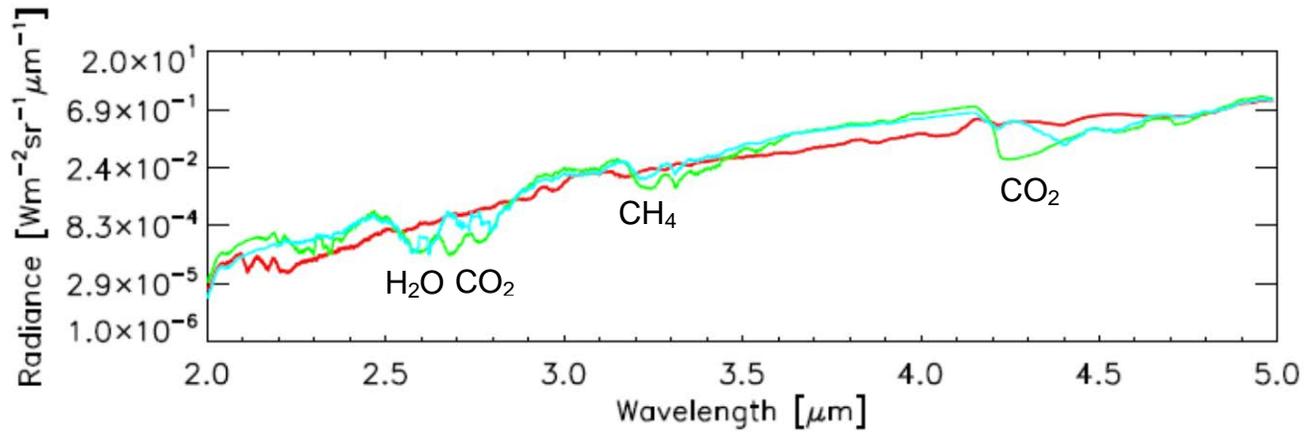

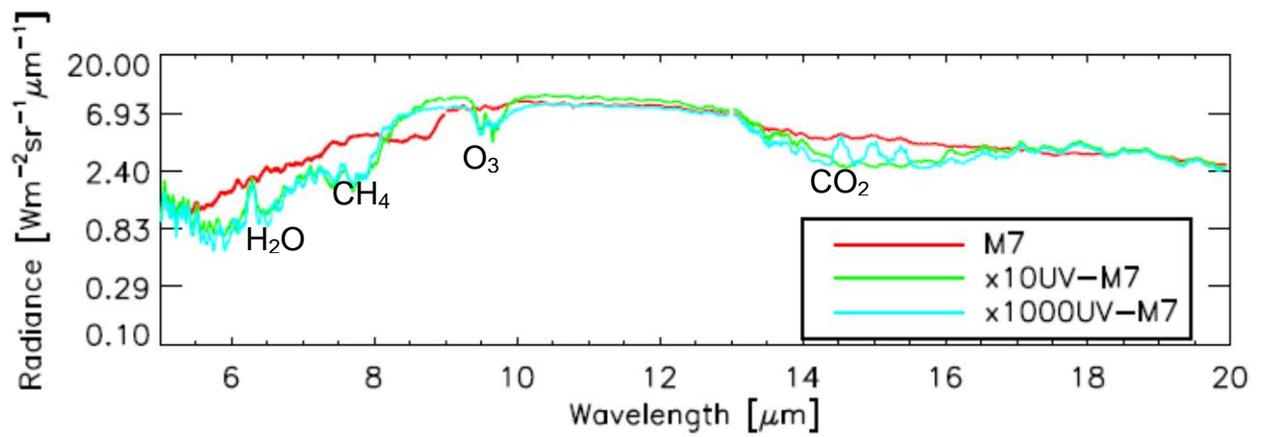

# Figure 12

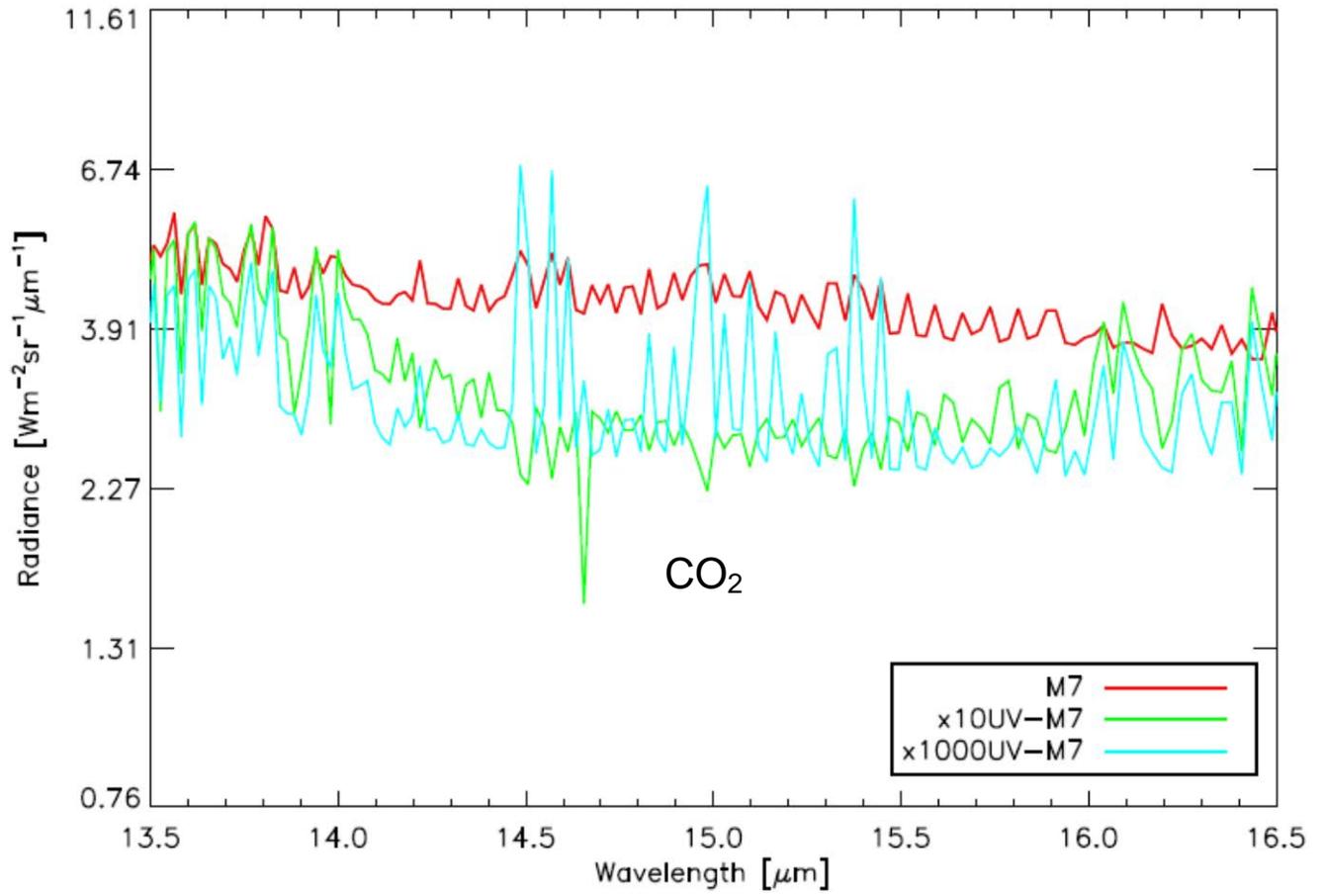

Figure 13

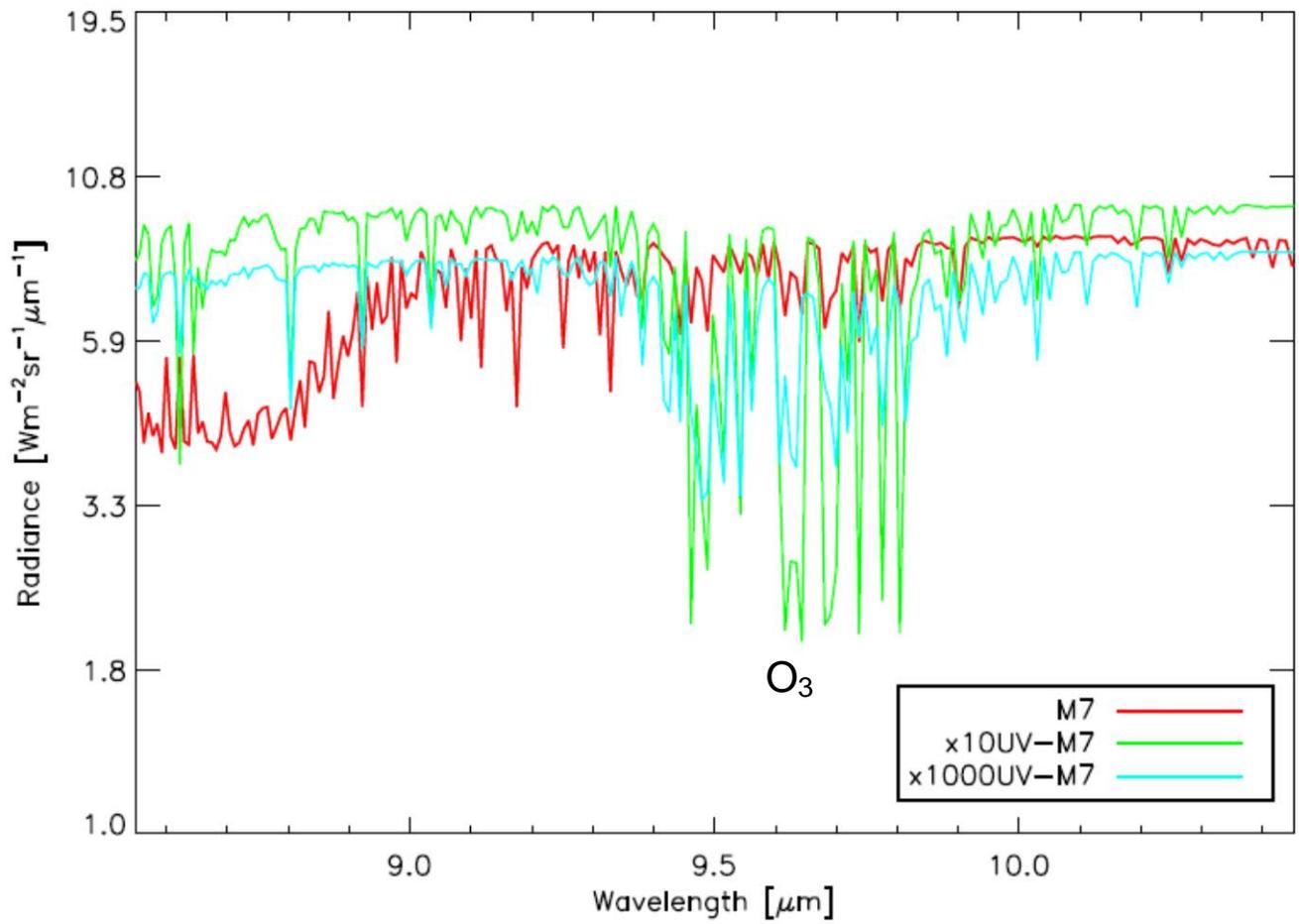

# Figure 14

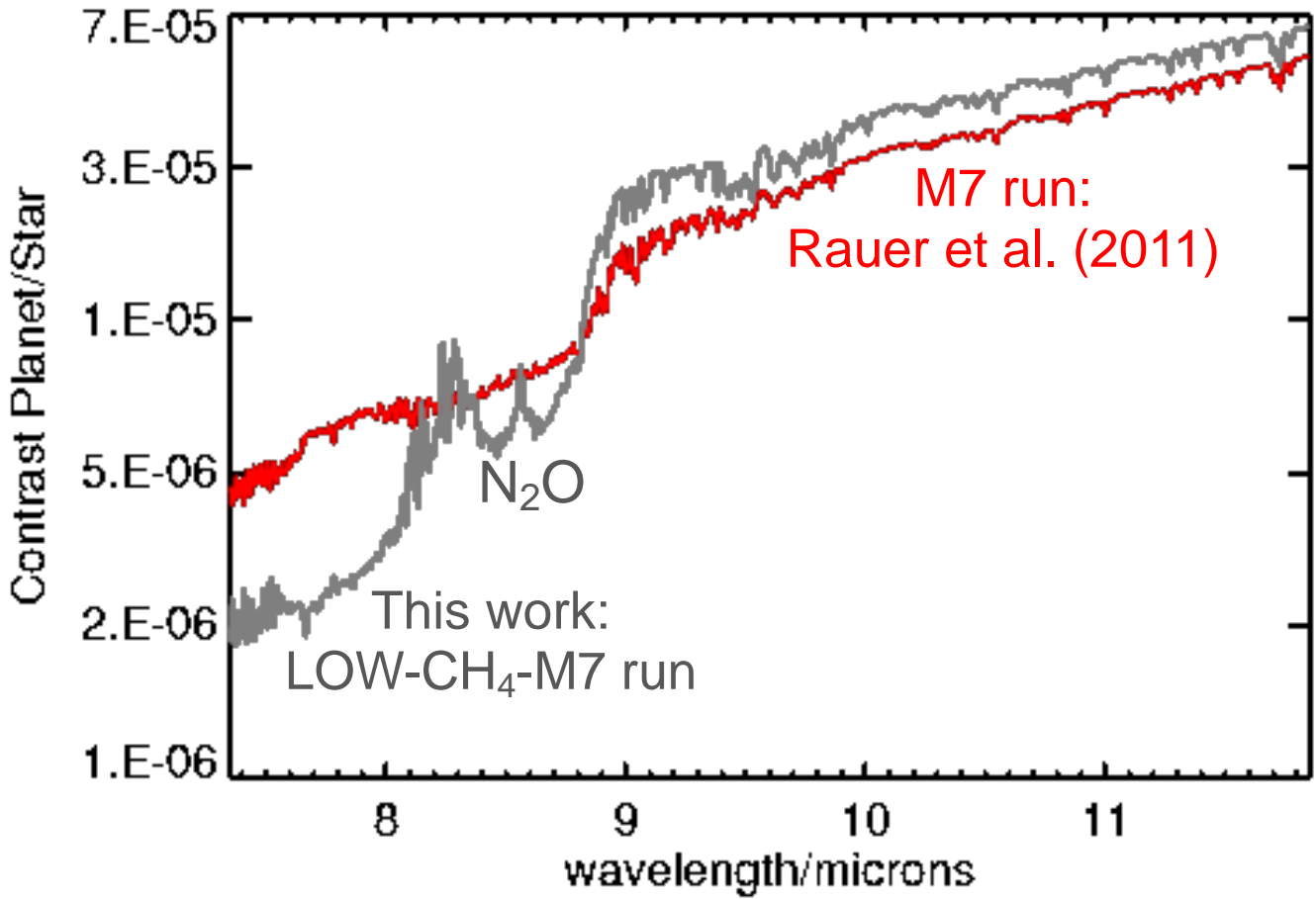